\documentstyle[referee]{mn}

\begin{document}

\title[Discs in Simulations: Comparison with Observations]
{Disc-like Objects in Hierarchical Hydrodynamical
  Simulations: Comparison with Observations}

\author[A. S\'aiz et al.]
{A. S\'aiz,$^1$ R. Dom\'{\i}nguez-Tenreiro,$^1$
  P.B. Tissera$^2$ and S. Courteau$^3$\\
  $^1$Departamento de F\'{\i}sica Te\'orica, C-XI. Universidad
  Aut\'onoma de Madrid, Madrid, E-28049, Spain\\
  $^2$I.A.F.E., Casilla de Correos  67, Suc.\ 28, Buenos
  Aires, 1428, Argentina\\
  $^3$University of British Columbia, Dept.\ of Physics and Astronomy,
  Vancouver, BC, Canada V6T 1Z1}

\maketitle

\begin{abstract}
  We present results from a careful and detailed analysis of the 
  structural and dynamical properties of a sample of 29 disc-like
  objects identified at $z=0$ in three AP3M-SPH fully consistent
  cosmological simulations.
  These simulations are realizations of a CDM hierarchical model, where
  an {\em inefficient\/} Schmidt law-like algorithm to model the
  stellar formation process has been implemented.
  We focus on properties that can be constrained with available
  data from observations of spiral galaxies, namely, the bulge
  and disc structural parameters and the rotation curves.
  Comparison with data from Broeils (1992), de Jong (1996)
  and Courteau (1996, 1997) gives satisfactory agreement,
  in contrast with previous findings using other codes.
  This suggests that the stellar formation implementation
  we have used has succeeded in forming compact bulges
  that stabilize disc-like structures in the violent
  phases of their assembly, while in the quiescent phases the gas has 
  cooled and collapsed according with the Fall \& Efstathiou standard model
  of disc formation.
\end{abstract}

\section{Introduction}
Understanding how galaxies form and evolve into the objects 
we observe today remains one of the most fundamental quest of
astrophysical research. 
Even if the field is still at its beginnings, 
the use of numerical approaches to study
how galaxies are assembled in a cosmological 
hierarchical scenario from primordial
fluctuations, seems promising.
The main advantage of these approaches (namely
hydrodynamical simulations) is that 
physics is introduced at a most general level, and the dynamical
processes relevant to galaxy assembly
(i.e., collapse, gas infall, interactions, mergers, instabilities...)
emerge naturally, rather than by assumption, and can be followed in detail.
Only the star forming processes need to be modelled.
These considerations emphasize the interest of hydrodynamical
simulations as an outstanding working tool to learn about
galaxy formation and evolution from primordial fluctuations.
For other approaches to this problem, see Manrique et al.\ (in preparation),
Firmani \& Avila-Reese (2000) and the references therein.

As discs are naturally produced in the collapse of a dissipative system
embedded in a gravitational field, learning how to get  well-behaved discs 
(i.e., similar to those observed in spirals) in such 
simulations is one important step towards 
understanding galaxy assembly in general.
According to the standard model of disc formation (Fall \& Efstathiou 1980),
disc-like structures with observable counterparts can form
provided that the diffuse gas  in the dark matter halo 
cools and collapses conserving its specific angular momentum, 
$\bmath j$.
In hierarchical scenarios, halo gas cooling occurs with
$\bmath j$ conservation  in the {\em quiescent\/}
phases of galaxy assembly, producing extended, thin discs with
an exponential mass density profile in the equatorial plane (Dalcanton,
Spergel \& Summers 1997). But {\em violent\/} phases play also a 
fundamental role in these scenarios. During these phases, the system undergoes
interactions and merger events, that can destroy discs 
that had possibly formed in previous quiescent phases.

Thus  far, no hydrodynamical simulation of galaxy formation
in fully consistent hierarchical cosmological scenarios  
had provided extended discs, with structural and dynamical properties 
similar to those observed in spiral galaxies
(see, however, recent work by Sommer-Larsen \& Dolgov 1999).
The objects formed were either a)~gaseous and too concentrated, 
when no stellar formation processes are considered at all
(Navarro \& Benz 1991; Navarro, Frenk \& White 1995;
Evrard, Summers \& Davis 1994;  Vedel, Hellsten \& Sommer-Larsen 1994;
Navarro \& Steinmetz 1997; 
Sommer-Larsen, Gelato \& Vedel 1999; 
Weil, Eke \& Efstathiou 1998 and references therein),
or b)~rather stellar and without discs, 
when these processes have been included 
(Navarro \& Steinmetz 2000; Thacker \& Couchman 2000).
In the former cases,  objects are too concentrated due to an excessive
$\bmath j$ loss in violent events; this is the so-called disc angular 
momentum catastrophe [DAMC].
In the latter cases, as discussed by
Tissera, S\'aiz \& Dom\'{\i}nguez-Tenreiro (in preparation, hereafter TSDT) 
the particular star formation implementations used
lead to a too early gas transformation into stars, leaving no gas to
cool and form discs at lower~$z$.  The effects of star formation
have also been considered by Katz (1992) and Steinmetz \& M\"uller (1995) who
obtained, in both cases, a three component system that resembles 
a spiral galaxy.  However, their simulations treat only the case
of a semi-cosmological modellization of the collapse of a
constant density perturbation in solid-body rotation.
Sommer-Larsen \& Dolgov (1999) have succeeded in avoiding excessive
$\bmath j$ loss in warm dark matter scenarios. Nevertheless,
their discs still do not match observations very well.

This inability to reproduce observed discs in fully consistent 
cosmological simulations,
represents clearly a major pitfall of present-day numerical simulations
as a working tool to learn about galaxy formation and evolution.
In a series of papers (Dom\'{\i}nguez-Tenreiro, Tissera \& S\'aiz 1998; S\'aiz,
Tissera \& Dom\'{\i}nguez-Tenreiro 1999; TSDT)
it has been shown that the 
object properties at the end of violent phases of evolution
critically depend on the structural stability of the disc-like
objects involved in interaction and merger events, on the one hand,
and on the availability of gas to regenerate a disc structure
after the last violent episode,
on the other hand. Concerning stability, cold thin discs are known
to be very unstable against non-axisymmetric modes
(Athanassoula \& Sellwood 1986; Binney \& Tremaine 1987;
Barnes \& Hernquist 1991, 1992; Martinet 1995 and references therein;
Mihos \& Hernquist 1994, 1996; Mo, Mao \& White 1998).
Massive haloes can stabilize discs, but not every halo is able
to stabilize the pure exponential disc that would form from
$\bmath j$-conserving diffuse gas collapse in its last quiescent phase.
In these cases, a central compact bulge is needed to ensure stability
(Christodoulou, Shlosman \& Tohline 1995; van den Bosch 1998, 2000).
Otherwise, non-axisymmetric instabilities could be easily triggered
by interactions and mergers, causing important $\bmath j$ losses
(that is, DAMCs), followed by strong gas inflow.

In that series of papers we also show that compact central bulges
form when star-forming processes are considered in
hierarchical hydrodynamical simulations, provided that
the particular star formation implementation used does not
lead to a too early gas exhaustion (i.e., {\em inefficient\/} star formation,
see Silk 1999). The physical processes
involved in disc formation and stability when an inefficient Schmidt
law-like stellar formation algorithm is implemented in these
simulations are discussed in detail in TSDT.

The aim of this paper is to show that 
discs identified in galaxy-like objects produced in these hierarchical
hydrodynamical simulations are, at a structural and dynamical level,
similar to those observed in spirals.
To this end, we analyze the structural and dynamical properties
of a sample of 29 disc-like objects formed in
three different such  simulations. We focus on those  
properties that can be constrained with available data for
observed spiral galaxies. 
These data are taken principally from the compilations
of galaxy structural parameters of Broeils (1992, hereafter B92),
de~Jong (1996) and Courteau (1996, 1997b, hereafter C97b). 
 
The paper is organized as follows:  $\S$\ref{simu} describes the simulations.
The general characteristics of disc-like objects at $z = 0$ are outlined in
$\S$\ref{objects}, and their bulge-disc decomposition and rotation
curves are discussed in $\S$\ref{budi} and $\S$\ref{rocu}, respectively.
Finally, a summary and conclusions are
presented in $\S$\ref{sumcon}.

\section{The Simulations}
\label{simu}
We consider three simulations, S1, S1b, and S1c%
\footnote{Simulation S1
  has been studied in Dom\'{\i}nguez-Tenreiro et al.\ (1998)
  and S\'aiz et al.\ (1999), and in
  Tissera \& Dom\'{\i}nguez-Tenreiro (1998) (their I.2 simulation).},
from different realizations of a given model for the primordial spectrum of the
density fluctuation field.
In each case, we follow the evolution of $64^3$ gas plus dark matter 
particles in a periodic box of comoving side
10 Mpc ($H_0 = 100 \: h$ km~s$^{-1}$ Mpc$^{-1}$ with $h = 0.5$)
using a SPH code coupled to
a high resolution AP3M code  (Thomas \& Couchman 1992; see 
Tissera, Lambas \& Abadi 1997 for details on SPH algorithm implementation).
The initial distributions of positions and velocities of the $64^3$ particles
are consistent with a standard CDM cosmology, with $\Omega_{\rm b} = 0.1$,
$\Lambda = 0$ and $b = 2.5$;
one out of ten of them is randomly chosen to be a gas particle. 
These initial distributions
are set by means of the ACTION algorithm, kindly provided
by E. Bertschinger. All, dark, gas and  
stellar particles have the same  mass, $2.6 \times 10^8$ M$_{\odot}$.
AP3M-SPH integrations
were carried out from $z = 10$ to the present using fixed time steps 
$\Delta t = 1.3 \times 10^7$ years.
The gravitational softening length is 3~kpc and the minimum allowed
smoothing length is 1.5~kpc. 
These simulations include a star formation prescription 
described by Tissera et al.\ (1997; see Tissera 2000
for a discussion on the resulting star formation history).
The gas cools due to radiative cooling.  We use the approximation for
the cooling function by Dalgarno \& McCray (1972) for a primordial 
mixture of hydrogen and helium.
Gas particles are transformed into stars if they are cold
($T_*< 3\times 10^{4}$ K),
dense ($\rho > \rho_{\rm crit}\approx 7\times 10^{-26}$
g~cm$^{-3}$), in a convergent flow and
satisfy Jean's instability criterion.
When a gas particle satisfies these conditions, it is
transformed into a star particle after a time interval ($\tau$)
over which its gas mass is being converted into stars according to
\begin{equation}
  \frac{{\rm d}\rho_{\rm gas}}{{\rm d}t} =
  -\frac{{\rm d}\rho_{\rm stars}}{{\rm d}t} = -c \frac{\rho_{\rm gas}}{t_{*}},
\end{equation}
where $c$ is the star formation efficiency,
$t_{*}$ is a characteristic time-scale
assumed to be proportional to the dynamical time of the particle 
($t_{*} \propto \rho^{-1/2}_{\rm gas}$) and
$\tau = \ln 100 \; t_{*} / c$ is the time interval over which
99 per cent of the gas mass in a particle  is expected to be transformed
into stars (Navarro \& White 1993).
We adopt a low star formation [SF] efficiency with $c = 0.01$
(for a discussion of the inefficient SF implementation,
see TSDT).
No supernovae explosion effects have been considered.
However, this low $c$ value  avoids the quick depletion of gas at high~$z$,
leaving enough remanent gas to form disc-like structures at low~$z$.
To some extent, this could mimic the effects of energy injection 
from supernovae explosions. The treatment of supernova feedback in
numerical simulations is still an open and complex  problem that remains
to be properly solved.

\section{Simulated Objects at $\lowercase{z}=0$}
\label{objects}
Galaxy-like objects are identified at their virial radius, $r_{200}$,
at $z=0$.
Dark matter haloes formed in the S1 simulation have been extensively studied
in Tissera \& Dom\'{\i}nguez-Tenreiro (1998, hereafter TDT98).
These haloes, as well as those 
formed in S1b and S1c simulations, are 
resolved with a relatively high number of particles
(see the number of dark mass particles, $N_{\rm dark}$,
per halo, in Table~\ref{General}).

The main baryonic objects embedded in these dark haloes can be identified 
by applying a friends-of-friends ({\sc fof}) algorithm to the baryonic
particle distribution, with a linking length of 10 per cent the mean total 
interparticle separation in the box.
Baryonic objects identified in the simulation at $z=0$ are galaxy-like
objects (GLOs) that span a range of morphological characteristics:
disc-like objects (hereafter, DLOs), spheroids and irregular objects.
DLOs have external extended, populated discs,
consisting mostly of gas, central stellar bulge-like concentration and,
in some cases, stars in a thick disc.
Spheroids are mostly composed of stars forming a relaxed, spheroidal
configuration, with a small amount of gas forming a small central disc.
Irregular objects do not have a defined morphology, and in most cases
they are the end product of a recent merger.
In this paper only DLOs will be analyzed.
Only those GLOs with a
total baryon number
$N_{\rm baryon}=
N_{\rm star} + N_{\rm gas}$
larger than  $N_{\rm baryon}^{\rm lim} = 150$ 
have been considered%
\footnote{For isolated GLOs, $N_{\rm baryon}$ is defined 
  unambiguously as 
  $N_{\rm baryon} = N_{\rm baryon}^{\sc fof}$, where 
  $N_{\rm baryon}^{\sc fof}$ is the baryon number as given by
  the friends-of-friends algorithm for a given GLO. However,
  some GLOs have close satellites that are included in the 
  friends-of-friends algorithm, even if these satellites are
  dynamically and structurally  different entities from 
  the GLO itself. 
  In this case, the actual GLO baryon number is
  $N_{\rm baryon} = N_{\rm baryon}^{\sc fof} - 
  \sum_{i} N_{\rm baryon}^{{\rm sat}, i}$, with  
  $N_{\rm baryon}^{{\rm sat}, i}$
  the number of baryonic particles in the $i$-th satellite entering
  in $N_{\rm baryon}^{\sc fof}$.}.
This criterion has been chosen as a compromise: taking 
$N_{\rm baryon}^{\rm lim}$  much higher than 150 would result into too few DLOs
in our sample; taking $N_{\rm baryon}^{\rm lim}$  much
lower than 150 would result into DLOs composed of too few particles to 
be properly analyzed.

Out of our 3 simulations, we are able to
identify different numbers of discs (9, 13 and 7, for a total 
sample of 29 objects) and spheroids or irregular objects
(5, 7 and 11, not studied in this paper) 
following the previous mass-limit criterion. 
DLOs are labelled by the three coordinates  (in 10$\times$mesh units)
that identify the cell
where the centre of mass
of their 
host halo is found in the total box.  
When several haloes (hosting either DLOs, spheroids or irregulars) 
are found in a given cell, a Greek letter is added to distinguish among them. 
Some haloes host more than  one baryonic object;
asterisks in Table~\ref{General}
identify haloes that include more than
one baryonic object (either DLO, spheroid or irregular).
In this case, the different  
components are distinguished by means of a capital 
letter in order of decreasing mass. For instance, DLO \#165C (in S1b) belongs
to a massive halo, where the more massive objects \#165A (spheroid)
and \#165B (irregular) are also found;
DLOs \#233A and \#233B (in S1), on the one side, 
and \#643$\beta$A and  \#643$\beta$B (in S1b), on the other side,
share the same halo.

The  virial radius, $r_{200}$, maximum limiting radius,
$R_{\rm max}$ (see  $\S$\ref{budi}),  dark, star and gas
particle number, $N_{\rm dark}$, $N_{\rm star}$ and $N_{\rm gas}$,
respectively, within spheres of radii $r_{200}$ and $R_{\rm max}$ for
each of the 29 DLOs of the sample are listed in Table~\ref{General}.

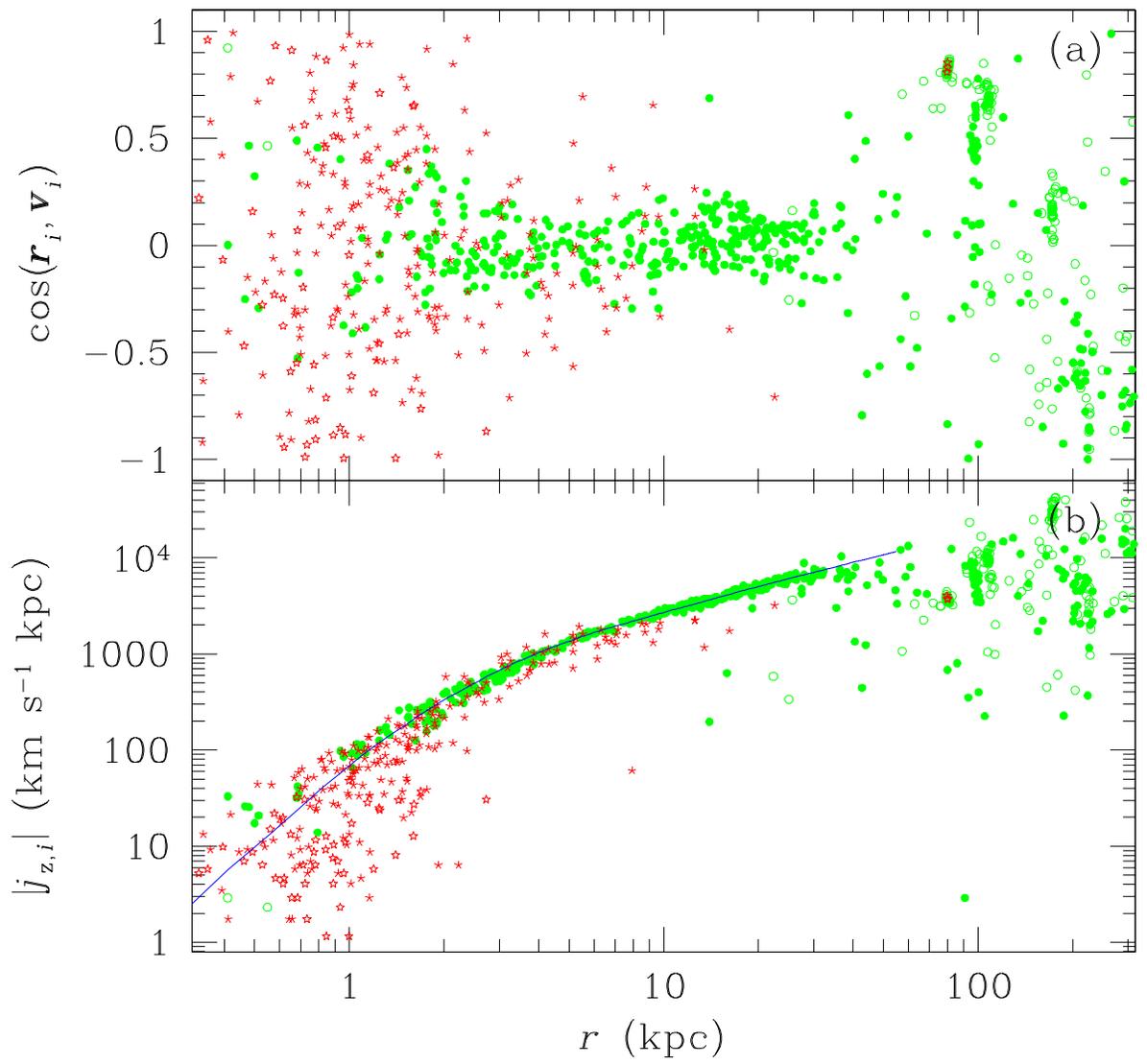
\begin{figure}
  \vspace{2cm}

  \caption{
    Upper (a):
    The cosine of the angle formed by the  position and velocity vectors
    of each baryon particle within halo \#242, versus their positions.
    {\em Dots\/}: gas particles, {\em stars\/}: stellar particles;
    {\em open symbols\/}: counter-rotating particles.
    Lower (b):
    Specific angular momentum component along $\bmath j_{\rm dis}$
    for each baryon particle of halo \#242, versus their positions.
    {\em Solid line\/}: $V_{\rm cir}(R) R$.
    }
  \label{DLOS}
\end{figure}

In Fig.~\ref{DLOS}(a)
we plot, for the $i$-th baryonic particle inside the halo
hosting DLO \#242, the cosine of the angle formed by its position
and velocity vectors, $\bmath r_i$ and $\bmath v_i$,
versus $r_i$, its distance to the DLO mass centre.
In Fig.~\ref{DLOS}(b) we plot $|j_{{\rm z},i}|$ versus $r_i$, where
$j_{{\rm z},i}$ is the component of the angular momentum per unit
mass of the $i$-th particle parallel to $\bmath j_{\rm dis}$ (disc specific
total angular momentum). The solid line is $V_{\rm cir}(R) R$, where
$V_{\rm cir}(R)$ is the circular velocity at projected distance $R$
(see $\S$\ref{rocu}). In these Figures, stars stand for stellar particles,
circles for gas particles and open symbols for counter-rotating particles
(i.e., with $j_{{\rm z},i}<0$) of any kind. These Figures show that most
gas particles placed at $r_i \la 30$~kpc move coherently along circular
trajectories (so that they have $\cos (\bmath r_i,\bmath v_i) \simeq 0$,
see Fig.~\ref{DLOS}(a)) on the equatorial plane (and so they verify
$\cos (\bmath j_i, \bmath j_{\rm dis}) \simeq 1$), that is, they have
$j_{{\rm z},i} \simeq |\bmath j_i| \simeq V_{\rm cir}(R_i) R_i$,
with a small dispersion around this value (see Fig.~\ref{DLOS}(b);
$R_i \simeq r_i$ if the $i$-th particle is on the equatorial plane):
the particles form a {\em cold thin disc\/}. In contrast with disc gas
particles, gas particles placed at $r_i \ga 30$~kpc (hereafter, halo gas
particles) do not show any order: roughly half of them are in
counter-rotation (open circles),
their $\cos (\bmath r_i,\bmath v_i) \in [-1,1]$
and their $|j_{{\rm z},i}|$ takes any value under the solid line.

Concerning stars, we see that most of them are found at $r_i \la 2$~kpc, where
they form a central compact relaxed structure, i.e., a bulge, with
$\bmath j_i$ without any preferred direction and very low $|j_{{\rm z},i}|$;
in some DLOs, as Figs.~\ref{DLOS}(a)~and~\ref{DLOS}(b) illustrate,
some stars are placed at $r_i \ga 2$~kpc in some orderly fashion,
forming a sort of thick disc.
Baryons forming other DLOs in our sample follow similar behaviour patterns.

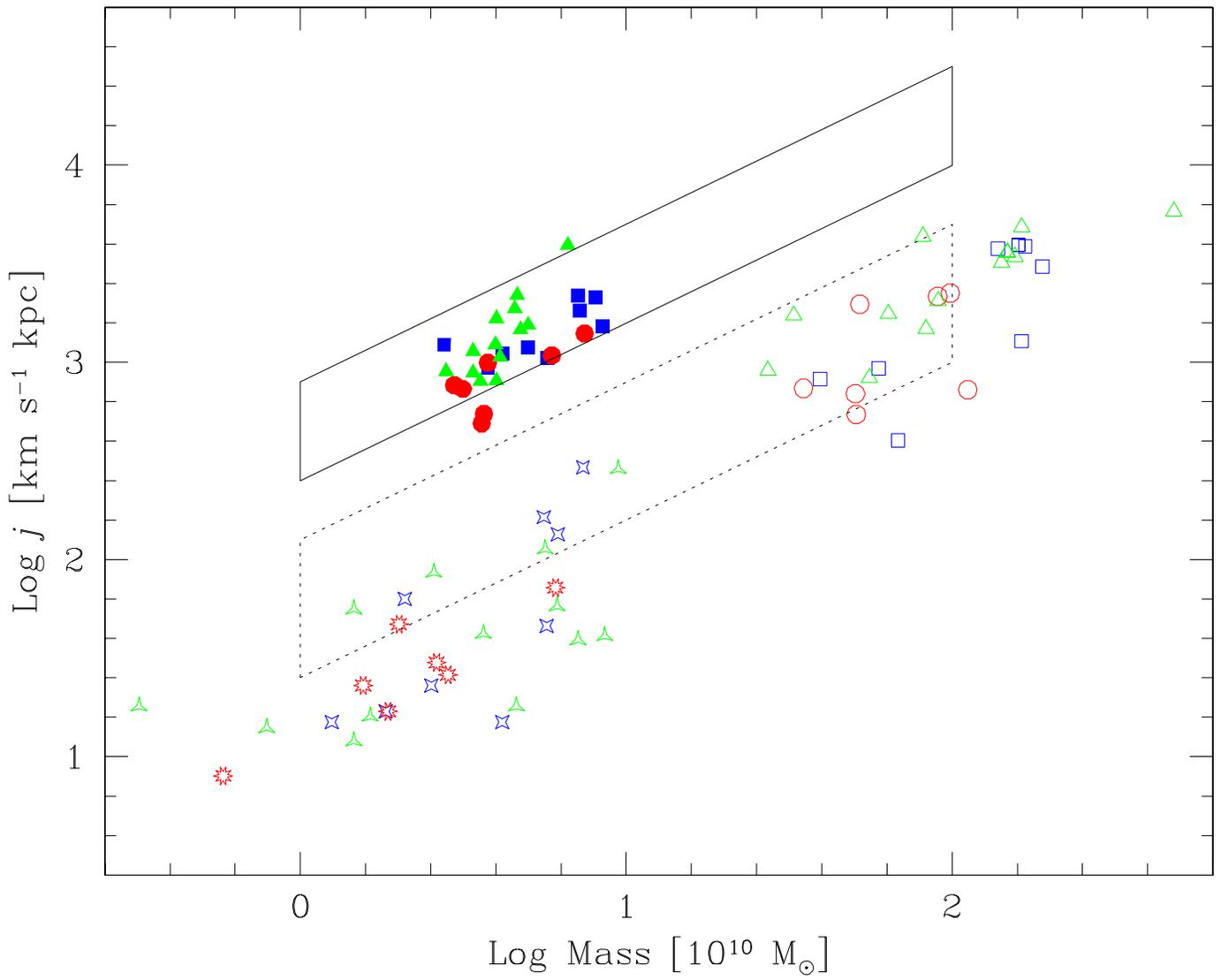
\begin{figure}
  \vspace{2cm}

  \caption{
    Specific angular momentum at $z=0$ versus the mass for the haloes
    (open symbols), for the inner 83 per cent of the gas component in the DLOs
    (filled symbols), and for their stellar component (starred symbols).
    Squares, triangles and circles correspond to DLOs formed in
    S1, S1b and S1c simulations, respectively. The solid (dotted) box
    encloses the region
    occupied by the spiral discs (ellipticals),
    as given by Fall (1983).
    }
  \label{JMvsM}
\end{figure}

In Fig.~\ref{JMvsM} we plot the specific total angular momentum at $z=0$
versus mass for dark haloes, $j_{\rm dh}$ (open symbols), for the inner
83 per cent of the disc gas mass (i.e., the mass fraction enclosed by
$R_{\rm opt} \equiv 3.2R_{\rm d}$ in a purely exponential disc, where
$R_{\rm d}$ is the disc scale length, see $\S$\ref{budi}), $j_{\rm g}$
(filled symbols), and for the stellar component of the DLOs in our
simulations, $j_{\rm s}$ (starred symbols).
We see that $j_{\rm g}$ is of the same order as $j_{\rm dh}$, so that
these gas particles have collapsed conserving, on average, their angular
momentum. Moreover, DLOs formed in our simulations are inside the box
defined by observed spiral discs in this plot (Fall 1983). In contrast,
$j_{\rm s}$ is much smaller than either $j_{\rm dh}$ or $j_{\rm g}$,
meaning that the stellar component in the central parts has formed out
of gas that had lost an important amount of its angular momentum.

A detailed description of the physical processes leading to DLO
configuration at $z=0$ we have just described is given in
Dom\'{\i}nguez-Tenreiro et al.\ 
(1998) and TSDT.
Here these processes are only briefly summarized:

The net effect of shocks and cooling in the quiescent phases of evolution
on disordered halo gas particles, is that they force them to a coherent
rotation with global specific angular momentum conservation.
Consequently,  in the quiescent phases of the evolution,
gas particles tend to settle at the inner halo regions,
and if the gravitational potential has some centre of axial (or central)
symmetry at these regions, gas particles will move on a plane on circular
orbits around this centre, forming a cold thin disc.
Most stars in the central bulges have been formed out of 
gas particles that had been involved in an inflow event with high $\bmath j$
loss; these compact central bulges create a gravitational potential that
is spherically symmetric on scales of a few kpc, ensuring $\bmath j$ 
conservation. This stabilizes the
system against gas inflows in future violent phases of the evolution.
In fact, in the next major merger event, those gas particles 
involved in the merger and with high $\bmath j$
will be  placed  on an intermediate disc; in 
their turn,
the stellar bulge of the smaller DLO involved
in the event is eventually destroyed by strong tidal forces; then,
incomplete orbital angular momentum loss puts most of its stars at some
distance of the centre, where, after some relaxation, they form a kind
of thick disc. Finally, most of the disc external gas particles are supplied
by infall of particles, either belonging to  baryonic  clumps (satellites) 
or  diffuse  component, completing the DLO assembly.
Because of the SF implementation used, gas in discs are not 
transformed into stars.
Then, simulated discs are mostly composed of gas. Observed discs in spirals
are mostly stellar, but the stellar discs inherit the structural and
dynamical characteristics of the gaseous discs out of which disc stars
have formed, so that, for the sake of comparison,
 the value of the parameters describing those properties
can be safely determined from the simulated gaseous discs.
A proper implementation of star formation
with supernovae feedback should lead to a self-regulated
star formation that would allow the formation of both compact stellar
bulges and thin stellar discs.

\section{Bulge-Disc Decomposition}
\label{budi}
Observationally,  the structural parameters of spirals are obtained from
a fit to the shape of the surface {\em brightness\/} profile. 
The galaxy luminosity profile is often decomposed into a bulge (b) and
disc (d) component, such as
\begin{equation}
  \Sigma^{\rm light}(R) = \Sigma^{\rm light}_{\rm b} (R)
  + \Sigma^{\rm light}_{\rm d} (R). 
  \label{surdes}
\end{equation}

Each individual profile can be modelled using the parametrization
from S\'ersic (1968), 
\begin{equation}
  \Sigma^{\rm light}_{\rm c} (R) = 
  \Sigma^{\rm light}_{\rm c} (0) \exp \left[ - \left({ R\over R_{\rm c}}
    \right)^{1/n} \right], \ {\rm{for} \hspace{0.5cm} c = {\rm b, d}},
  \label{surprof}
\end{equation}
and where 
$n$ is a shape parameter taken to be $n = 1$ for the disc component 
(i.e., a pure exponential law) and that is left as a free parameter 
for the bulge component. 
The de~Vaucouleurs profile for spheroids 
corresponds to a choice of $n = 4$. 
$\Sigma^{\rm light}_{\rm c} (0)$ and $R_{\rm c}$ are
the central surface brightness 
and scale length for the bulge  and disc components. 

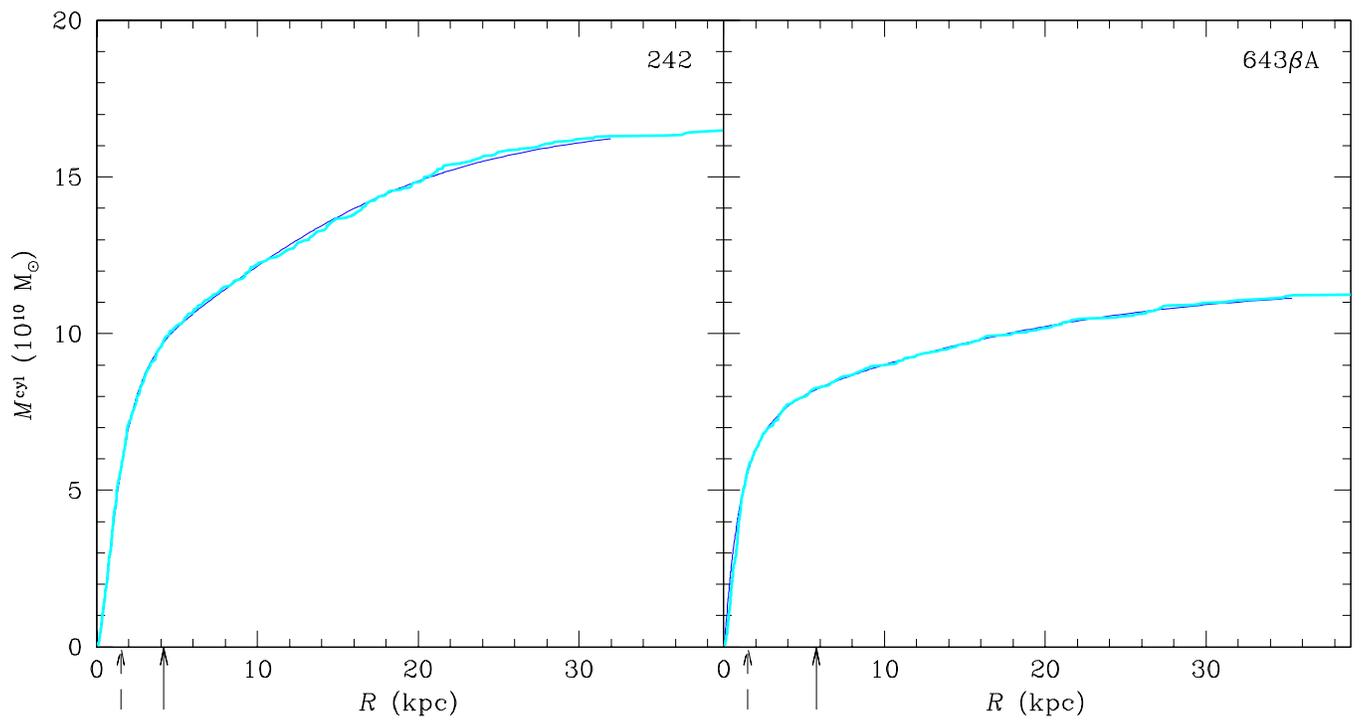
\begin{figure}
  \vspace{2cm}

  \caption{
    {\em Thick lines\/}: the histograms for the 
    integrated projected mass density of baryons, $M^{\rm cyl} (R_i)$,
    for DLOs \#242 and \#643$\beta$A  (see Table~\ref{General}).
    {\em Thin lines\/} are their best fits by means
    of a bulge-disc decomposition. A solid-line arrow marks $R_{\rm trans}$
    and a dashed-line arrow marks $R_{\rm min}$.
    }
  \label{MCyl}
\end{figure}

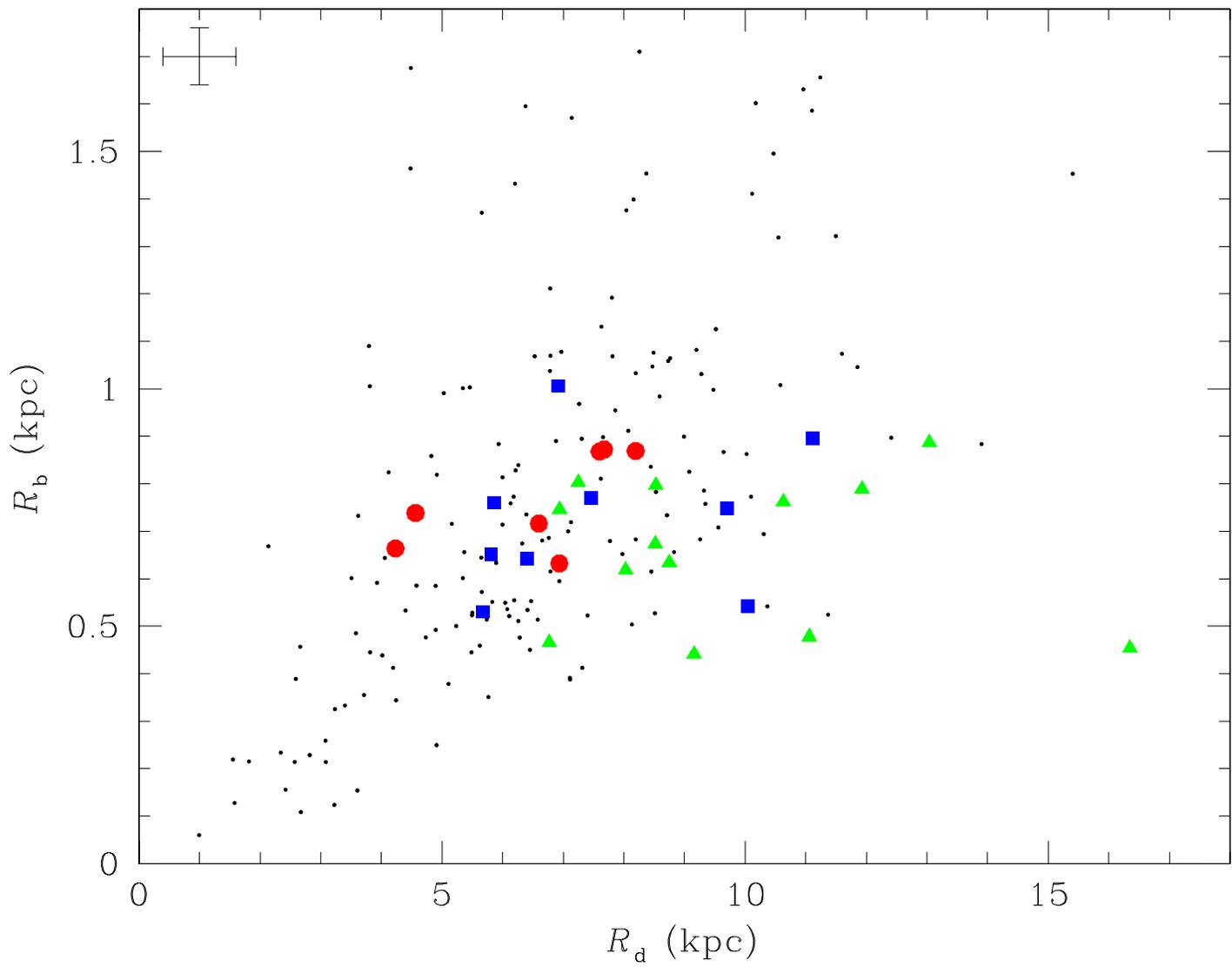
\begin{figure}
  \vspace{2cm}

  \caption{
    The bulge and disc scale lengths, $R_{\rm b}$ and $R_{\rm d}$, for
    double exponential fits. Squares, triangles and circles correspond
    to DLOs formed in S1, S1b and S1c simulations, respectively.
    Dots are from 1D decompositions of surface brightness profiles given
    by CdJB96 and Courteau (1997a). The error bars in the top left corner
    are the typical dispersions in the $R_{\rm b}$ and $R_{\rm d}$ values
    for DLOs, obtained from Monte Carlo realizations of their profiles.
    }
  \label{RbRd}
\end{figure}

The same profile form has been fit to the projected {\em mass\/} 
density of baryons, $\Sigma^{\rm mass} (R)$, for the DLOs in our sample.
$\Sigma^{\rm mass} (R)$ is calculated by averaging concentric rings
centred at the centre of mass of each DLO. 
Projections on the plane normal to the total angular momentum of
each DLO (disc plane) have been used.
These projected densities are binning dependent and  
somewhat noisy. To circumvent these problems, the {\em integrated 
  projected\/} mass density in concentric cylinders of radius $R$
and  mass $M^{\rm cyl}(R)$, 
\begin{equation}
  M^{\rm cyl}(R) = 2 \pi \int_0^{R} \Sigma^{\rm mass} (R') R' {\rm d}R', 
  \label{Mcil}
\end{equation}
has been used as fitting function, instead of the projected mass density 
itself.
Cumulative histograms $M^{\rm cyl}(R_i)$  
are constructed by adding equal height steps at those positions $R_i$
where baryon particles are placed.
Fig.~\ref{MCyl} shows these $M^{\rm cyl} (R_i)$ histograms for
two DLOs.
Simulated discs, as well as discs observed in spirals,
do not have a sharp end.
We have taken as maximum limiting radius for the simulated discs
$R_{\rm max} = 4.5 R_{\rm d}$%
\footnote{A recent analysis by Pohlen,
  Dettmar \& L\"utticke (2000) suggests a shallower truncation radius,
  with $R_{\rm max} \sim 3 R_{\rm d}$.} 
which corresponds to a limiting
or truncation radius in surface brightness of 
$B = 26.5$ mag arcsec$^{-2}$ assuming that discs have 
$B(0) \simeq 21.65$ mag arcsec$^{-2}$. 
As minimum limiting radius we have taken $R_{\rm min} = 1.5$~kpc%
\footnote{The fitting determination of bulge/disc parameters
  is stable against changes in $R_{\rm min}$. These changes affect 
  mainly the $R_{\rm b, eff}$ parameter values, that in the less favoured case,
  $R_{\rm min} = 0.0$~kpc, change on average only by less than a 10 per cent;
  $R_{\rm d}$ variations, on the other hand,
  are on average lower
  than the corresponding Monte Carlo errors due to finite sampling
  (see Fig.~\ref{RbRd}).}.
Because the total baryon number 
inside
$R_{\rm max}$ for
each object is known (see Table~\ref{General}; 
this is equivalent to knowing the {\em total\/} DLO mass), 
we are left with four free fitting parameters
among $n$, 
$\Sigma_{\rm b}(0)$, $R_{\rm b}$, $\Sigma_{\rm d}(0)$ and $R_{\rm d}$.
An updated version of the {\sc minuit} software from CERN library
has been used to make the fits.

To compare the size of bulges with different shape (i.e., with different
bulge shape parameter $n$) the $R_{\rm b}$ parameter is meaningless
because its value, as determined from the fitting procedure, 
depends on $n$. In fact, when bulges with the same physical scale
and different shape are considered, $R_{\rm b}$
varies by orders of magnitude when $n$ varies by a factor of
some few units (see Eqs.~(\ref{ReffRb})~and~(\ref{ReffRb2}) below).
To circumvent this shortcoming, an {\em effective\/} bulge scale
length or half light radius, $R_{\rm b, eff}$, is often used by observers.
In our case, it  will be  defined  through the projected
bulge mass profile by the condition
$M^{\rm cyl}_{\rm b}(R_{\rm b, eff}) = M_{\rm b}/2$, where $M_{\rm b}$
is the bulge mass (i.e., it is a half mass radius).
$R_{\rm b, eff}$ is related to $R_{\rm b}$ through:
\begin{equation}
  \gamma\left(2 n, \left(\frac{R_{\rm b, eff}}{R_{\rm b}}\right)^{1/n}\right) =
  \frac{\Gamma\left(2 n\right)}{2} ,
  \label{ReffRb}
\end{equation}
where $\gamma\left(m,u\right)$ and $\Gamma\left(m\right)$
are the incomplete and complete gamma functions.
For $n = 1$, Eq.~(\ref{ReffRb}) gives $\frac{R_{\rm b, eff}}{R_{\rm b}} =
1.678$; for $n = 4$ one finds $\frac{R_{\rm b, eff}}{R_{\rm b}} = (7.669)^4$.
For arbitrary $n$, Eq.~(\ref{ReffRb}) can be approximated by
\begin{equation}
  \frac{R_{\rm b, eff}}{R_{\rm b}} \simeq (2.0 n - 0.324)^n .
  \label{ReffRb2}
\end{equation}

The results for the effective bulge scale length, $R_{\rm b, eff}$,
and disc scale length, $R_{\rm d}$, 
corresponding to the optimal
bulge shape parameter, $n_{\rm opt}$, and, also,
to $n=1$  (exponential bulge profile) are given in Table~\ref{BulDis}, 
together with their corresponding $\chi^2$ per degree of freedom parameter.
For comparison, Table~\ref{BulDis} also gives the disc scale length and
the $\chi^2$ per degree of freedom for a de~Vaucouleurs profile ($n=4$).   
Note that the  disc scale parameter, $R_{\rm d}$, is, in most cases, 
nearly independent of the bulge shape parameter 
(either $n_{\rm opt}$, $n = 1$ or $n = 4$).
Knowing the length and mass scale parameters for
simulated bulges and discs allows us to determine their
{\em mass\/} bulge-to-disc  ratios,
\begin{equation}
  \frac{B}{D} = \frac{\Gamma(2n + 1)}{2} 
  \left(\frac{R_{\rm b}}{R_{\rm d}}\right)^2
  \frac{\Sigma^{\rm mass}_{\rm b}(0)}{\Sigma^{\rm mass}_{\rm d}(0)}.
  \label{BD}
\end{equation}
These ratios are given in Table~\ref{BulDis}. This Table shows another 
interesting parameter characterizing DLO structure: the
{\em transition\/} radius, $R_{\rm trans}$, where 
$\Sigma^{\rm mass}_{\rm b} (R) = \Sigma^{\rm mass}_{\rm d} (R)$.

In Fig.~\ref{MCyl} we have drawn the best fits to the integrated 
projected baryon
mass density, $M^{\rm cyl} (R)$, for DLOs \#242 and \#643$\beta$A.
A solid line arrow marks $R_{\rm trans}$; a dashed line arrow
marks the lower limit of the fitting interval $R_{\rm min} = 1.5$~kpc.
This Figure illustrates the good quality achieved in the fittings,
even for the regions where $R < R_{\rm min}$, and hence the stability
of the bulge/disc parameter determination against changes in 
$R_{\rm min}$, as mentioned before.

We see in Table~\ref{BulDis}
that bulges for most DLOs are more adequately modelled with an
exponential profile ($n=1$) than with a de~Vaucouleurs profile ($n=4$). 
This agrees with observational results for early and late-type 
spiral galaxies (e.g.\ Andredakis, Peletier \& Balcells 1995;
Courteau, de Jong \& Broeils 1996 [CdJB96];
Moriondo, Giovanelli \& Haynes 1999 [MGH99]). 
Furthermore,  Andredakis et al.\ (1995) 
and CdJB96 have found that
the bulge shape parameter is correlated with spiral type.
In particular, CdJB96 find that most Sb--Sc and all high surface 
brightness spiral galaxies of later type are best fitted with a
$n = 1$ profile, most Sa--Sab galaxies are best modelled with a $n=2$
bulge, and only a small fraction of late type spirals have bulges
that follow the de~Vaucouleurs law.
This distribution of shape parameters, $n_{\rm opt}$, 
is most similar to that of DLOs in Table~\ref{BulDis}.
Based on the grid of shape parameters in Andredakis et al.\ (1995)
and CdJB96,
most of our simulated DLOs seem to be of type $T \geq 1$ or greater
(see $n_{\rm opt}$ in Table~\ref{BulDis}). 
The  dependence of the observed $n$ values on 
the morphological type suggests a difference
in the formation and evolution of the bulge component. 

The values of the $R_{\rm b}$ and $R_{\rm d}$ parameters,
corresponding to a double exponential fit, are plotted in
Fig.~\ref{RbRd}  for all the DLOs in our sample; 
squares, triangles and circles correspond
to DLOs formed in S1, S1b and S1c simulations, respectively.
The error bars in the top left corner of the Figure are the typical 
dispersions in the $R_{\rm b}$ (vertical) and $R_{\rm d}$ (horizontal)
parameters values, arising from finite sampling 
and  calculated through Monte Carlo
realizations of the DLO profiles,
as given by  Eqs.~(\ref{surdes})~and~(\ref{surprof}),
with $N_{\rm baryon}$ particles and $n = 1$.
For comparison, in Fig.~\ref{RbRd} we also 
show the $R_{\rm b}$ and $R_{\rm d}$ values given
by CdJB96 and Courteau (1997a) from one-dimensional 
luminosity decompositions of Sb--Sc spirals (points).
We get   average bulge/disc scale length ratios of
$R_{\rm b} / R_{\rm d} = 0.09 \pm 0.03$ for DLOs in S1,
$R_{\rm b} / R_{\rm d} = 0.07 \pm 0.03$ for DLOs in S1b and
$R_{\rm b} / R_{\rm d} = 0.12 \pm 0.03$ for DLOs in S1c.
The  ratio for the whole sample
of 29  DLOs is $R_{\rm b} / R_{\rm d} = 0.09 \pm 0.04$,
which  compares well with the values found by CdJB96 in
the $r$-band, 0.13 $\pm$ 0.07, and de~Jong (1996) at $K$ alone,
0.09 $\pm$ 0.04. 
MGH99 also find $R_{\rm b} / R_{\rm d} = 0.11 \pm 0.06$. 
In all cases, consistency between available data and our DLO sample 
is good.  
Note that we miss DLOs with small values of $R_{\rm b}$ and/or $R_{\rm d}$ 
as compared with the range of values quoted by these authors. 
It simply reflects our lack 
of dynamical resolution to resolve smaller  structures.

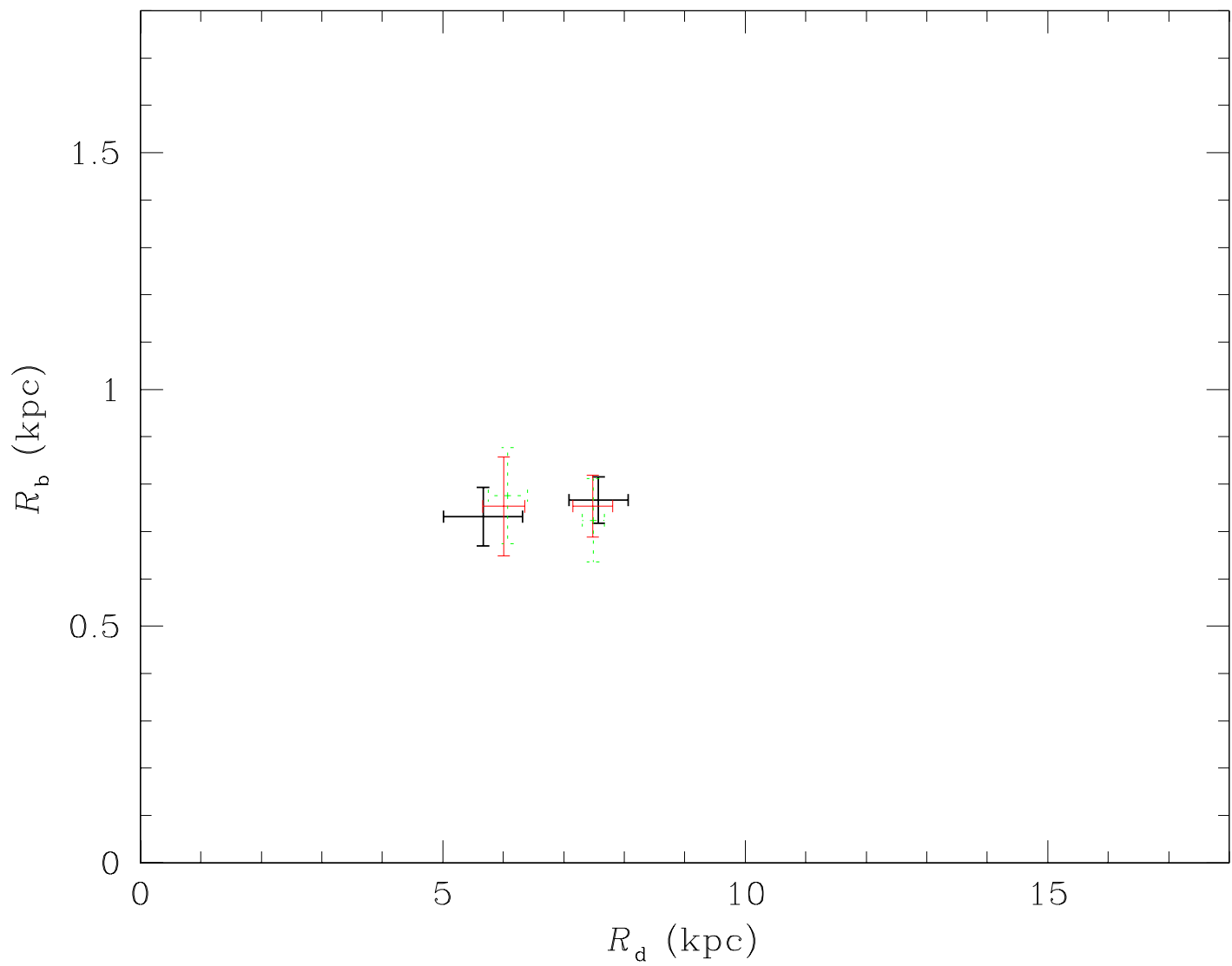
\begin{figure}
  \vspace{2cm}

  \caption{
    Average values of the $R_{\rm b}$ and $R_{\rm d}$ parameters and their
    dispersions, obtained from Monte Carlo realizations of DLOs \#242 and \#545
    $\Sigma^{\rm light}(R)$ profiles (see text). {\em Thick lines\/}
    correspond to $\frac{\Upsilon_{\rm b}}{\Upsilon_{\rm d}}=1$ (that is,
    $\Sigma^{\rm light}(R)\propto\Sigma^{\rm mass}(R)$), {\em thin lines\/}
    to $\frac{\Upsilon_{\rm b}}{\Upsilon_{\rm d}}=3$ and {\em dotted lines\/}
    to $\frac{\Upsilon_{\rm b}}{\Upsilon_{\rm d}}=5$.
    }
  \label{RbRdTest-Exp}
\end{figure}

In Fig.~\ref{RbRd} we compare scale lengths obtained from one-dimensional 
surface {\em brightness\/} decompositions with scales obtained from
projected {\em mass\/} density decompositions with bulge shape parameter
$n = 1$. As compared with the disc, the bulge is typically less
prominent in surface brightness than in projected mass density,
since its mass-to-light ratio  ($\Upsilon_{\rm b}$)
is likely to be higher than the disc mass-to-light ratio
($\Upsilon_{\rm d}$). It is not clear, a priori, if this fact 
makes the previous comparisons meaningless.
To find out if this is indeed the case, 
one would have  to compare the values of the
$R_{\rm b}$ and $R_{\rm d}$ parameters as
obtained from a fit to the surface brightness profile  of a given DLO with
those obtained from a fit to its   projected mass density profile.
Since this is not possible, in order to assess the feasibility of the previous comparisons,
we adopt the following  method:

Assuming that both $\Upsilon_{\rm b}$ and $\Upsilon_{\rm d}$ do not 
depend on $R$, a given DLO with
double exponential $\Sigma^{\rm mass}(R)$ profile 
characterized by values of $R_{\rm b}$, $R_{\rm d}$, 
$\Sigma^{\rm mass}_{\rm b}(0)$ and $\Sigma^{\rm mass}_{\rm d}(0)$,
would have a surface brightness profile,
$\Sigma^{\rm light}(R)$, given by Eqs.~(\ref{surdes})~and~(\ref{surprof})
with the same  $R_{\rm b}$ and $R_{\rm d}$ parameters,
but with $\Sigma^{\rm light}_{\rm b}(0) = \frac{\Sigma^{\rm mass}_{\rm b}(0)}
{\Upsilon_{\rm b}}$ and 
$\Sigma^{\rm light}_{\rm d}(0) = \frac{\Sigma^{\rm mass}_{\rm d}(0)}
{\Upsilon_{\rm d}}$.
The previous comparisons will make sense if fittings to Monte Carlo
realizations of a given $\Sigma^{\rm mass}(R)$ profile
and to Monte Carlo realizations of its surface brightness
counterpart, $\Sigma^{\rm light}(R)$, with the same number of particles, 
lead to the same values of the $R_{\rm b}$ and $R_{\rm d}$ parameters.
This test has been performed for $\Sigma^{\rm mass}(R)$ corresponding
to  DLOs \#242 and \#545 (from S1) in the sample. Results are shown
in Fig.~\ref{RbRdTest-Exp} for $\frac{\Upsilon_{\rm b}}{\Upsilon_{\rm d}}$
= 1, 3 and 5, from where  
we see that  consistent values of $R_{\rm b}$ and 
$R_{\rm d}$ are obtained (within errors resulting from finite DLO sampling).
Therefore, the mass and length structural parameters are decoupled, 
making the comparisons valid.

This compatibility between the simulated mass scale lengths and 
observed luminosity  scale lengths in the $r$ and $K$ bands, 
suggests that the projected light density (i.e., stars) follows the 
projected mass density.  This could be expected if the star formation
rate in a disc were proportional to the baryon surface mass density
(see Kennicutt 1998 and Silk 1999 for a discussion).

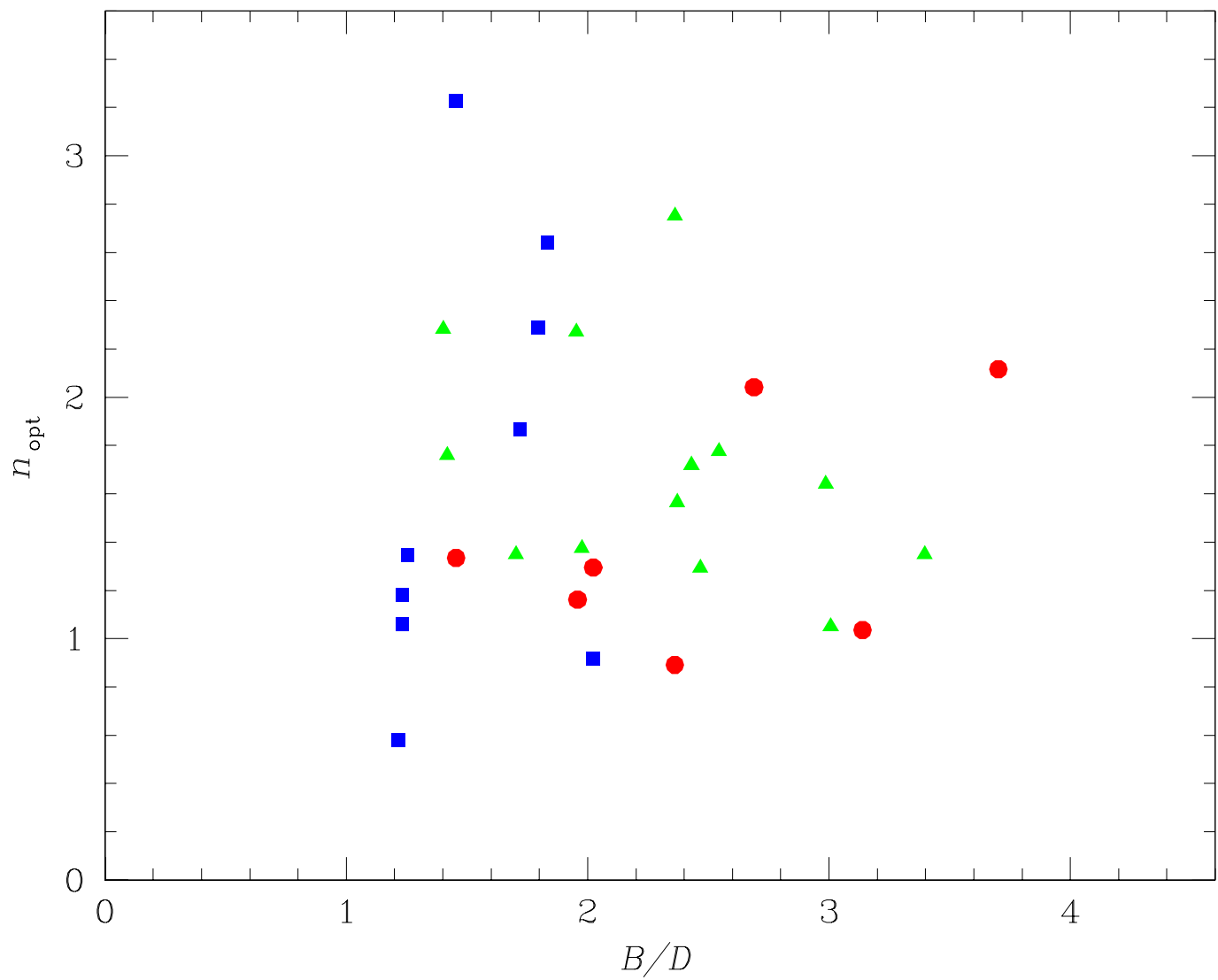
\begin{figure}
  \vspace{2cm}

  \caption{
    The mass bulge-to-disc ratios $B/D$ versus the optimal bulge shape
    parameter, $n_{\rm opt}$, for the DLO sample.
    Squares, triangles and circles correspond
    to DLOs formed in S1, S1b and S1c simulations, respectively.
    }
  \label{NOPTBD}
\end{figure}

The {\em mass\/} $B/D$ ratios given in Table~\ref{BulDis} are plotted in 
Fig.~\ref{NOPTBD} versus the corresponding $n_{\rm opt}$.  Note that no
correlation is observed 
between $B/D$ and $n_{\rm opt}$.
These {\em mass\/} $B/D$ ratios are considerably higher than the
corresponding {\em luminosity\/} $B/D$ ratios of late-type systems.
But it is difficult to decide at the present whether  
the central baryon concentrations found in DLOs are excessive or not,
as both, disc and bulge, mass-to-light ratios, 
would be required to compare with observed {\em luminosity\/} $B/D$ ratios.
In fact, the only disc galaxy where the masses of the disc and
bulge/central component can be estimated directly is the Milky Way.
Sommer-Larsen \& Dolgov (1999) obtain $B/D \simeq$ 0.2 -- 0.4,
for a disc scale length of 2.5 -- 3.0 kpc.
For other galaxies, some estimations give 
$\Upsilon_{\rm d} \approx$ 1--2  (in solar units,
for $h = 0.65$; Courteau \& Rix 1999, hereafter CR99).
Values in these ranges would
make  the {\em mass\/} $B/D$ ratios obtained in the simulations
roughly compatible with {\em luminosity\/}
$B/D$ ratios from observations, but more  accurate measures of
$\Upsilon_{\rm b}$ and $\Upsilon_{\rm d}$ are required before a 
conclusion can be reached.
It is worthwhile to note that decompositions of extended \hbox{H\,{\sc i}}
rotation curves (RCs) of some galaxies from Broeils'
compilation (B92), using 
sub-maximal discs (Bottema 1993, 1997; CR99), 
yield {\em mass\/} $B/D$ ratios of about~1 (see Rhee 1996).
Also, MGH99 find that 5 out of 23 Sa--Scd
galaxies for which they obtain a reliable fit to the rotation curve (RC)
for both the bulge and the disc component, have {\em mass\/} $B/D$ ratios
of about~1 or larger.  Note, however, that the majority of these
decompositions are ill-constrained, as different combinations 
of bulge, disc, and halo profiles yield equally good agreement with the
data (van Albada et al.\ 1985, CR99).

Current observational uncertainties on $\Upsilon_{\rm b}$ and 
$\Upsilon_{\rm d}$ preclude any consistency check for our derived
baryon mass distributions.  For more stringent constraints, one 
must turn instead to dynamical observables such as RCs.

\section{DLO Rotation Curves}
\label{rocu}
The shapes  of spiral RCs are determined by their three
dimensional total (i.e., bulge, disc and dark matter halo) mass distribution.

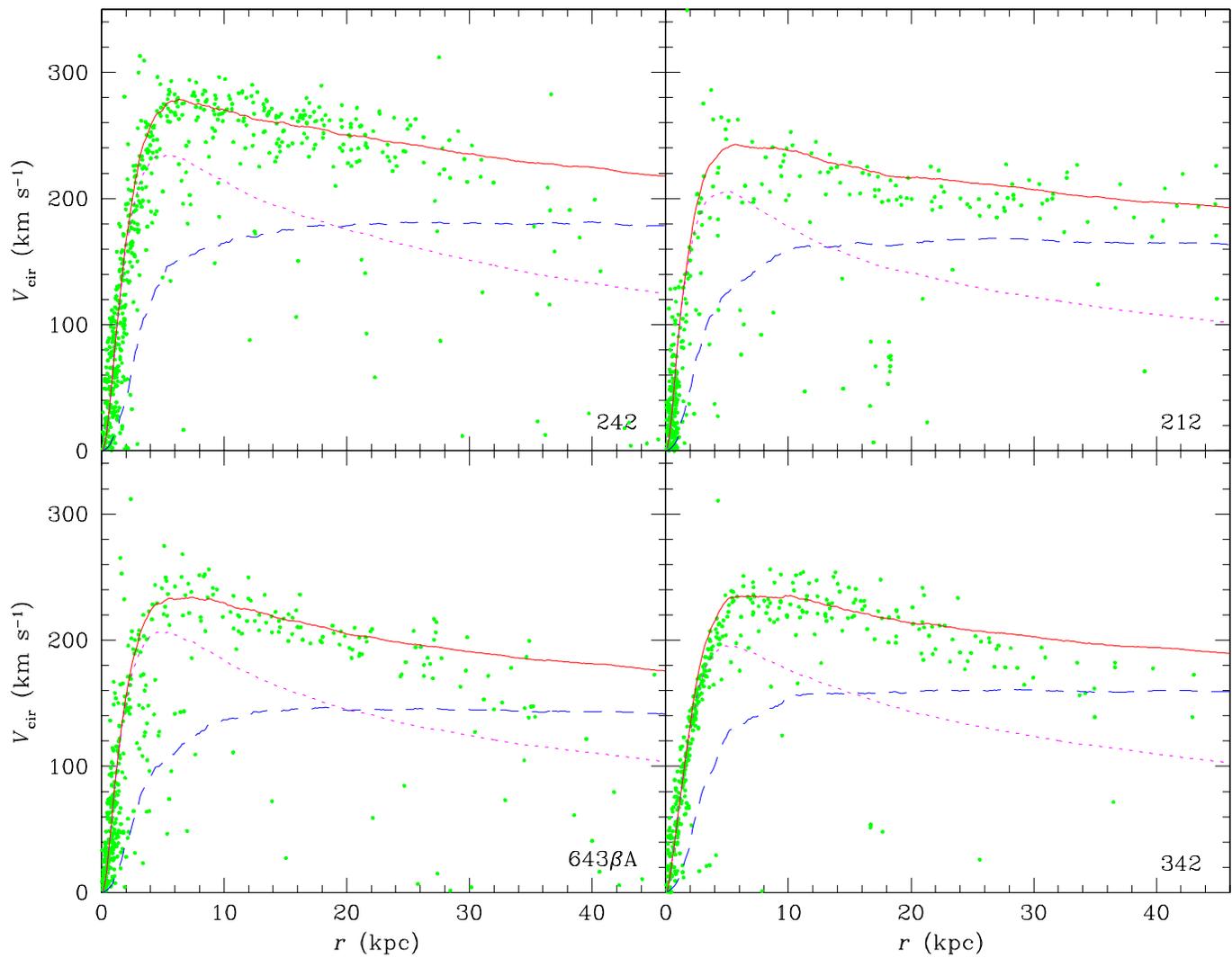
\begin{figure}
  \vspace{2cm}

  \caption{
    {\em Points\/}: the rotational
    velocity distributions for the baryon particles of
    DLOs \#242 of S1, \#212 and \#643$\beta$A of S1b and \#342 of S1c
    (see text).
    {\em Solid lines\/}:
    circular velocity curves (see 
    Eqs.~(\ref{eq:velcir})~and~(\ref{eq:velcirII}))
    for the total mass distribution; {\em dotted lines\/}: the baryonic 
    contribution $V_{\rm bar}(r)$, and {\em dashed lines\/}: the dark halo 
    contribution $V_{\rm dm}(r)$. 
    }
  \label{RotCur}
\end{figure}

The RCs of the simulated objects have been constructed by adding up
in quadrature the contributions to circular rotation from the bulge,
the disc (both of them  formed by baryons, either gas or stars, 
see $\S$\ref{objects}), so that
$V_{\rm bar}^2 (r) = V_{\rm b}^2 (r) + V_{\rm d}^2 (r)$,
and the dark matter halo, $V_{\rm dm} (r)$, 
\begin{equation}
  V_{\rm cir}^2 (r) = V_{\rm bar}^2 (r) + V_{\rm dm}^2 (r).
  \label{eq:velcir}
\end{equation}
We have adopted a softened Plummer potential (Evrard et al.\ 1994) 
\begin{equation}
  V_{\rm i}^2 (r) = {G M_{\rm i} (< r) r^2 \over
    \left(r^2 + \epsilon_{\rm g}^2\right)^{3/2}},
  \hspace{2cm} {\rm i = bar, dm}
  \label{eq:velcirII}
\end{equation}
where $M_{\rm bar}(< r)$ and $M_{\rm dm} (< r)$ are
the total baryonic and dark matter masses,
respectively,   inside a sphere or radius $r$.
The adaptive AP3M scheme produces an effective dynamical gravitational 
softening at $z=0$ of $\epsilon_{\rm g} = 3$~kpc 
at the central region of the DLOs.
The RCs for several DLOs in our sample are given in Fig.~\ref{RotCur}.
Points are the tangential velocity component of the baryonic 
particles 
(projected on the disc plane), solid lines are $V_{\rm cir}(r)$, dashed 
lines are $V_{\rm dm}(r)$ and point lines $V_{\rm bar}(r)$. 
Note that the halo 
component, $V_{\rm dm}(r)$, gives rise to an almost {\em flat\/} contribution
to the RCs  beyond the very inner regions.
These RCs fit very well the tangential velocity component
of the baryonic particles up to $R \sim R_{\rm max}$,
suggesting that these particles are in rotationally supported 
equilibrium within the potential well produced by both the dark haloes
and the baryons themselves (TDT98).

On scales encompassing the baryonic object (i.e., $\leq R_{\rm max}$),
DLOs have RCs that are declining in their outer parts, and some of them
have central spikes produced by the bulge, similar to those found
in some observed galaxies with {\em mass\/} $B/D \simeq 1$ or larger
(for example, NGC~6674 and NGC~7331 in Rhee 1996, or UGC~89, UGC~1013 
and UGC~1238 in MGH99).
It is customary to try to explain the shapes of RCs as a combination
of various `fundamental plane' parameters such as
global size (i.e., a spatial and a mass scale; Rubin et al.\ 1985),
central mass concentration (Rhee \& van Albada 1996), 
and luminosity (Persic, Salucci \& Stel 1996).
Here, the RC shapes for simulated DLOs 
are parametrized   
as a function of disc scale length, $R_{\rm d}$, as spatial scale,
rotation velocity at $R_{2.2} = 2.2 \, R_{\rm d}$%
\footnote{$R_{2.2}$ is
  the radius at which a {\em purely\/} exponential disc
  would reach its maximum circular velocity.},
$V_{2.2} = V_{\rm cir}(R_{2.2})$,
as mass size,  and central mass concentration, via the 
maximum or peak rotation velocity, $V_{\rm cir}^{\rm peak}$, and 
the radius where this is reached, $R_{\rm peak}$   
(C97b).
Some combinations of these parameters have also been explored, such as
the logarithmic slope, $LS$ (Casertano \& van Gorkom 1991 [CvG91];  
B92; de Blok, McGaugh \& van der Hulst 1996).
The $LS$ is observationally defined as the slope of a straight line that
fits the RC in log -- log scale, from $R_{2.2}$ to $R_{\rm out}$ (this last
point is usually taken to be the last measured RC point, but see C97b
for variations on this operational procedure).
For simulated objects in our sample, the RCs are nearly linear
in log -- log scale for $R > R_{2.2}$, so that $LS$ does not appreciably
change for 
$R_{\rm out} > R_{2.2}$. 
The values of these parameters measured in simulated DLOs
have been compared with those
measured in RCs from: a)~Sb--Sc field spirals drawn from the 
Courteau sample, with long-slit H$\alpha$ spectra and $r$-band
photometry (see C97b), and b)~Sb--Sc spirals from Broeils' compilation (B92) 
of extended \hbox{H\,{\sc i}}
RCs with surface photometry.  We keep only galaxies
with $V_{2.2} > 150$ km~s$^{-1}$ (only one of Broeils' galaxies has 
$V_{2.2}$ in the range $\left[150, 180 \right]$ km~s$^{-1}$).

Apart from these parameters characterizing the shapes of RCs at scales
of the baryonic objects, other interesting parameters
at halo scales (i.e., $\sim r_{200}$), are the velocity at the virial radius
$V_{200} = V_{\rm cir}(r_{200})$ and the spin parameter, 
$\lambda = \frac{J_{\rm tot} |E_{\rm tot}|^{1/2}}{G M_{\rm tot}^{5/2}}$,
with $J_{\rm tot}$, $E_{\rm tot}$ and $M_{\rm tot}$ the total angular momentum,
energy and mass for all particles inside $r_{200}$.
Note that both $V_{200}$ and $\lambda$ are not observable.

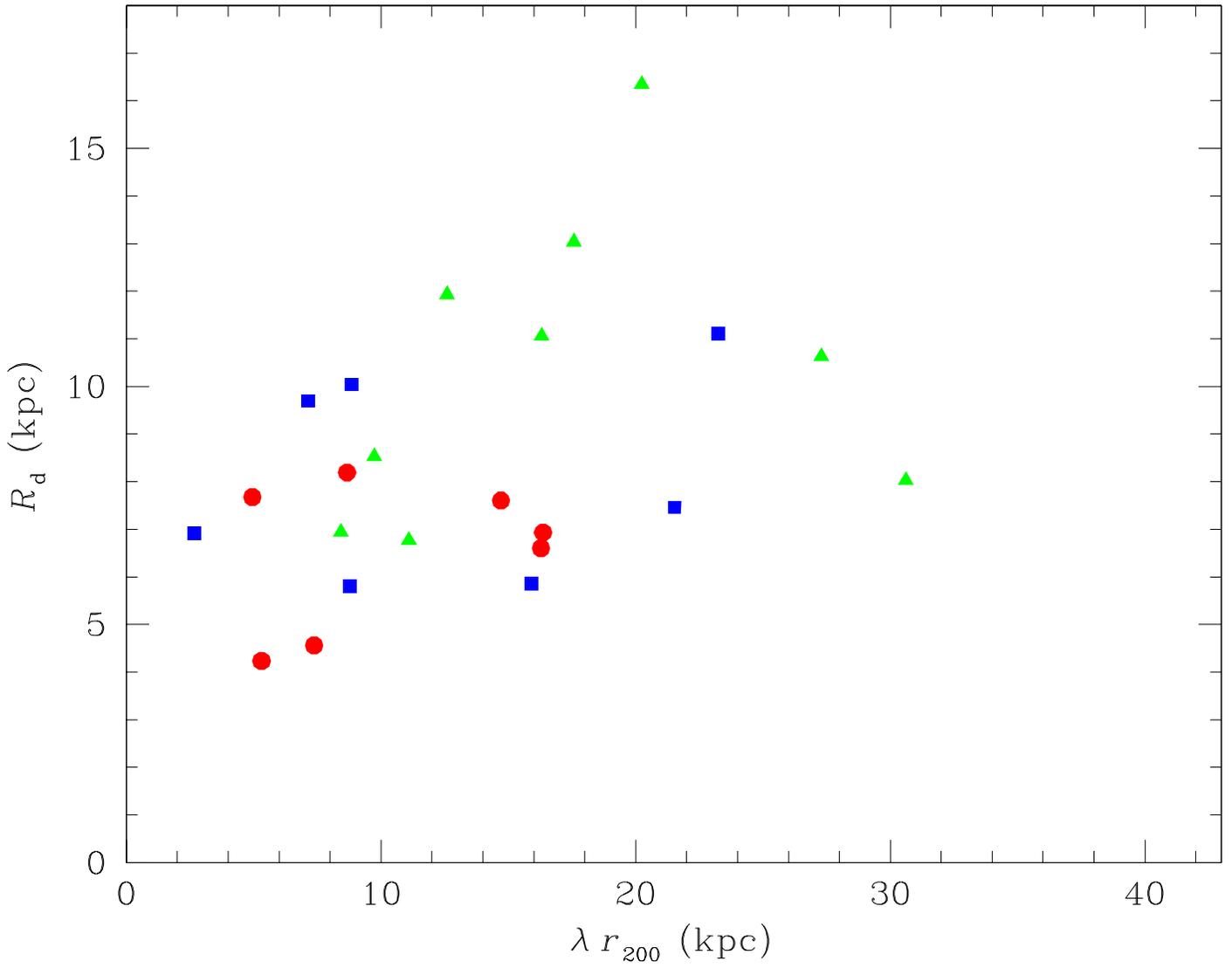
\begin{figure}
  \vspace{2cm}

  \caption{
    The disc scale length $R_{\rm d}$ versus $\lambda \, r_{200}$
    for the DLO sample. Symbols are as in Fig.~\ref{NOPTBD}.
    DLOs belonging to haloes that include several objects
    (either DLOs, spheroids or irregulars) have not been considered.
    }
  \label{LambRd}
\end{figure}

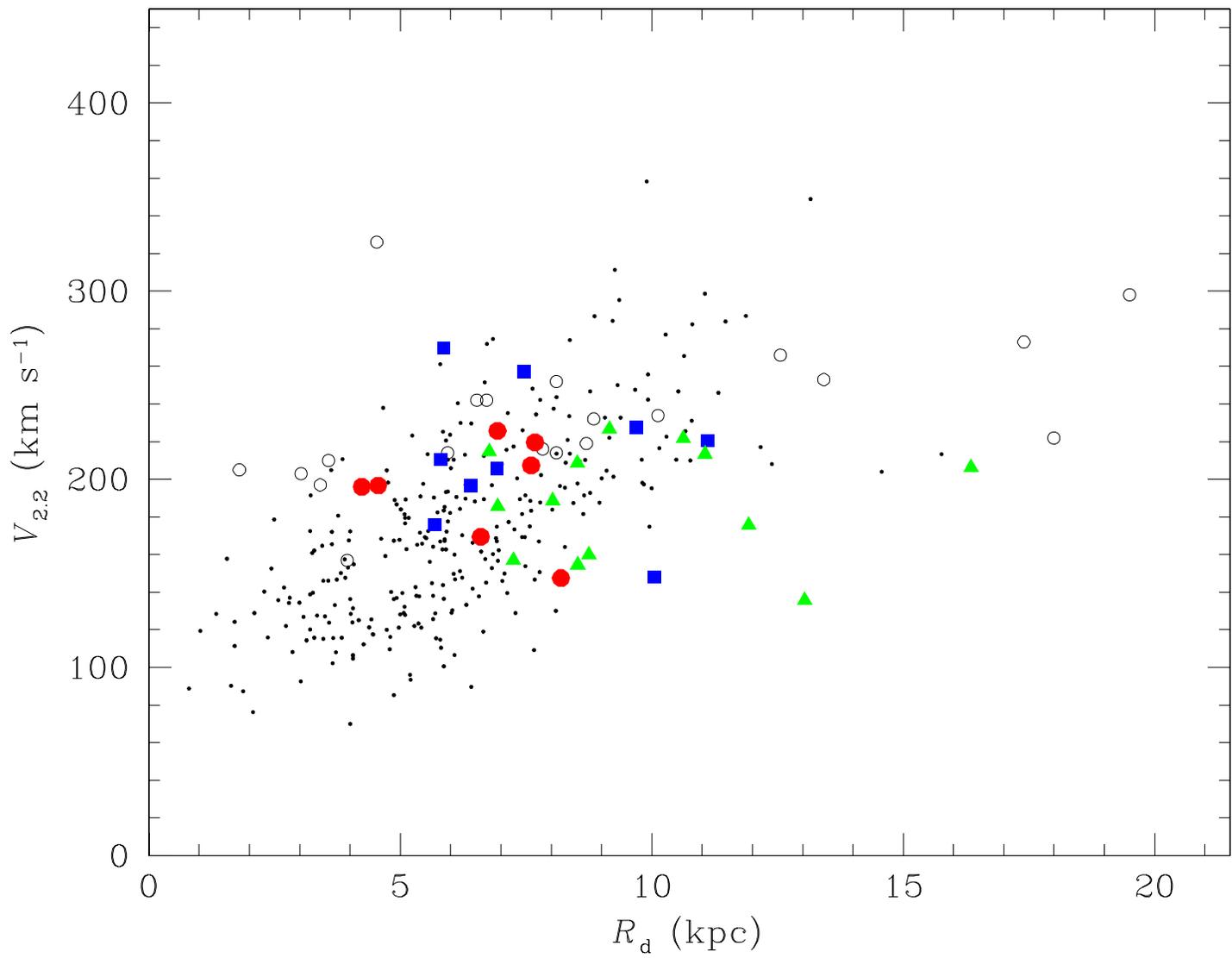
\begin{figure}
  \vspace{2cm}

  \caption{
    The $V_{2.2}$ velocities versus the disc scale lengths for the DLO sample.
    Squares, triangles and circles correspond
    to DLOs formed in S1, S1b and S1c simulations, respectively.
    Dots are data from the Courteau sample
    (H$\alpha$ spectra, see C97b). Open circles are data from spirals with 
    \hbox{H\,{\sc i}}
    rotation curves from CvG91 and Rhee (1996).
    }
  \label{RdV22}
\end{figure}

In Table~\ref{Dynamical} we give $V_{200}$, $\lambda$ and the RC parameters
for our DLO sample. Note
that the $\lambda$ values are within their expected range 
for a standard CDM scenario 
(Warren et al.\ 1992; Dalcanton et al.\ 1997; Lemson \& Kauffmann 1998).
In Fig.~\ref{LambRd} we plot $R_{\rm d}$ versus $\lambda \, r_{200}$. A
correlation appears in the sense that haloes with higher spin host more
extended discs. 
In the framework of semianalytical
models of quiescent {\em pure\/} (i.e., bulgeless)
disc formation with $\bmath j$ conservation
one approximately has $R_{\rm d} \propto \lambda \, r_{200}$
(see Dalcanton et al.\ 1997; Mo et al.\ 1998);
we find an important dispersion due to
the loss of angular momentum by {\em bulge\/} particles involved
in DAMCs in the violent phases of the evolution.
The DLO values of the  
parameters $R_{\rm d}$ and $V_{2.2}$ are
consistent with observations (see Tables~\ref{BulDis}~and~\ref{Dynamical} and
Figs.~\ref{RbRd}~and~\ref{RdV22}).
By virtue of selecting DLOs with $N_{\rm baryon} > 150$, 
we find high RC amplitudes mostly with $V_{2.2} > 150$ km~s$^{-1}$. 
Our simulated DLOs are either intermediate spirals 
(100 km~s$^{-1} < V_{2.2} < 180$ km~s$^{-1}$, 9 objects),
large bright spirals ($V_{2.2} > 180$ km~s$^{-1}$, 
$R_{\rm d} > 5.25$~kpc, 18 objects), or   
compact bright spirals ($V_{2.2} > 180$ km~s$^{-1}$, 
$R_{\rm d} < 5.25$~kpc, 2 objects).
Our sample does not include any dwarf galaxy. 
As expected, $V_{2.2}$ is correlated with
$V_{200}$, albeit with 
a dispersion. This dispersion is provided
by the particularities of the evolutionary history of each DLO in the
sample, namely, the number and characteristics of the interactions
and merger events involved in its assembly (see TDT98 for a discussion).

The mass distribution of real spirals and their decomposition into
bulge, disc and halo components cannot be inferred uniquely from 
observed RCs, as different decompositions yield equally good 
agreement with the data (van Albada et al.\ 1985;
Broeils \& Courteau 1997; CR99).
One has to postulate different scalings (mass-to-light ratios) between 
the bulge and disc  
components in order to infer the mass distribution
of the halo. 
Common constraints for (M/L)$_{\rm disc}$ include 
a)~the so-called `maximum disc (or maximal light) hypothesis'
(van Albada \& Sancisi 1986), which postulates that the halo dark 
mass component needed to fit the RCs should be minimum, or, 
equivalently, that the contribution of the disc and bulge to the 
inner parts of RCs should be maximum,  
and b)~sub-maximal discs with 
$\left( {V_{\rm lum} \over V_{\rm cir}} \right)_{R_{2.2}} \sim 0.60 \pm 0.10$. 
This constraint is obtained either by matching the vertical velocity 
dispersion, scale height and scale length of a thin exponential disc 
which yields the maximum disc rotation (Bottema 1993), or, independently,
showing that the Tully--Fisher relation of bright spirals is 
independent of surface brightness (CR99). 

A maximum disc is defined as
$\left( {V_{\rm lum} \over V_{\rm cir}} \right)_{R_{2.2}} 
= 0.85 \pm 0.10$, at the 95 per cent confidence level (Sackett 1997), 
corresponding to $\left( {M_{\rm lum} \over M_{\rm cir}} \right)_{R_{2.2}} 
= 0.72 \pm 0.17$. 
To the  sub-maximal discs described above corresponds a  
$\left( {M_{\rm lum} \over M_{\rm cir}}
\right)_{R_{2.2}}$ of $0.36 \pm 0.12$. 
One notes that this choice of sub-maximal disc corresponds to half 
the amount of luminous matter at $R_{2.2}$ compared to the maximum disc
case. 

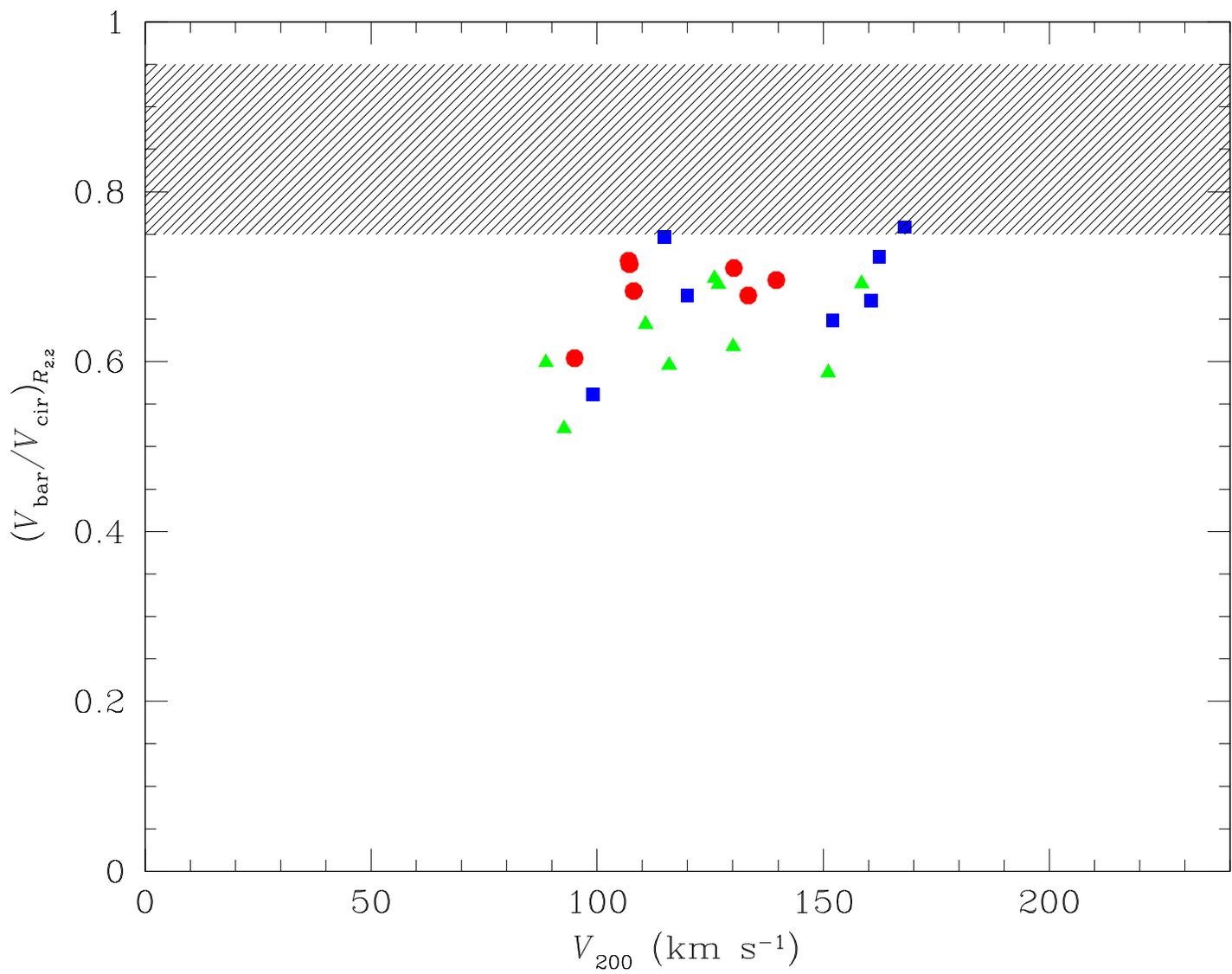
\begin{figure}
  \vspace{2cm}

  \caption{
    The $\left( {V_{\rm bar} \over V_{\rm cir}} \right)_{R_{2.2}}$ 
    versus $V_{200}$ for the DLO sample. Symbols are as in Fig.~\ref{NOPTBD}.
    DLOs belonging to haloes that include several objects
    (either DLOs, spheroids or irregulars) have not been considered.
    The shaded area shows the 95 per cent confidence interval 
    for maximum discs (see text).
    }
  \label{VRATIO}
\end{figure}

Let us now consider our simulations.
$V_{\rm lum,2.2} = V_{\rm lum}(R_{2.2})$ can be measured 
directly once $R_{\rm d}$ is determined for each DLO 
(assuming that $V_{\rm lum} = V_{\rm bar}$)
and the different hypotheses above can be tested.  
Results are given in Table~\ref{Dynamical} for both the optimal
and double exponential fits. $V_{\rm bar,2.2}$ is correlated with $V_{2.2}$
(and, also, with $V_{200}$, albeit with more dispersion).
In Fig.~\ref{VRATIO} we plot the 
$\left( {V_{\rm bar} \over V_{\rm cir}} \right)_{R_{2.2}}$ ratios
versus $V_{200}$ for DLOs in our sample (double exponential fits;
no substantial changes occur taking instead the values corresponding to
optimal fits).
The shaded area shows the 95 per cent interval for maximum discs according
to Sackett (1997).
Only DLOs \#126 and \#545 (from S1) are inside this area, as they
have $\left( {V_{\rm bar} \over V_{\rm cir}} \right)_{R_{2.2}} = $ 0.765
and 0.763, respectively.  These DLOs are somewhat peculiar:
DLO \#126 has a particularly populated disc; DLO \#545 has two satellites
close to it.
We see that most DLOs 
lie 
 inside the allowed interval for sub-maximal
discs. The mean value for the simulated DLOs is $\left( {V_{\rm bar} 
    \over V_{\rm cir}} \right)_{R_{2.2}} = 0.67 \pm 0.07$, in 
excellent agreement with Bottema (1993, 1997) and CR99.

In Fig.~\ref{VRATIO} only a slight correlation appears between
$\left( {V_{\rm bar} \over V_{\rm cir}} \right)_{R_{2.2}}$ and $V_{200}$
with an important dispersion, in the sense that less massive haloes
have slightly lower values of
$\left( {V_{\rm bar} \over V_{\rm cir}} \right)_{R_{2.2}}$.
This presumably reflects the fact that
less massive haloes are somewhat more centrally concentrated than
more massive ones (Navarro, Frenk \& White 1995b; see also TDT98).
The dispersion, on the other hand, reflects the particularities of the
assembly of each DLO in the sample.

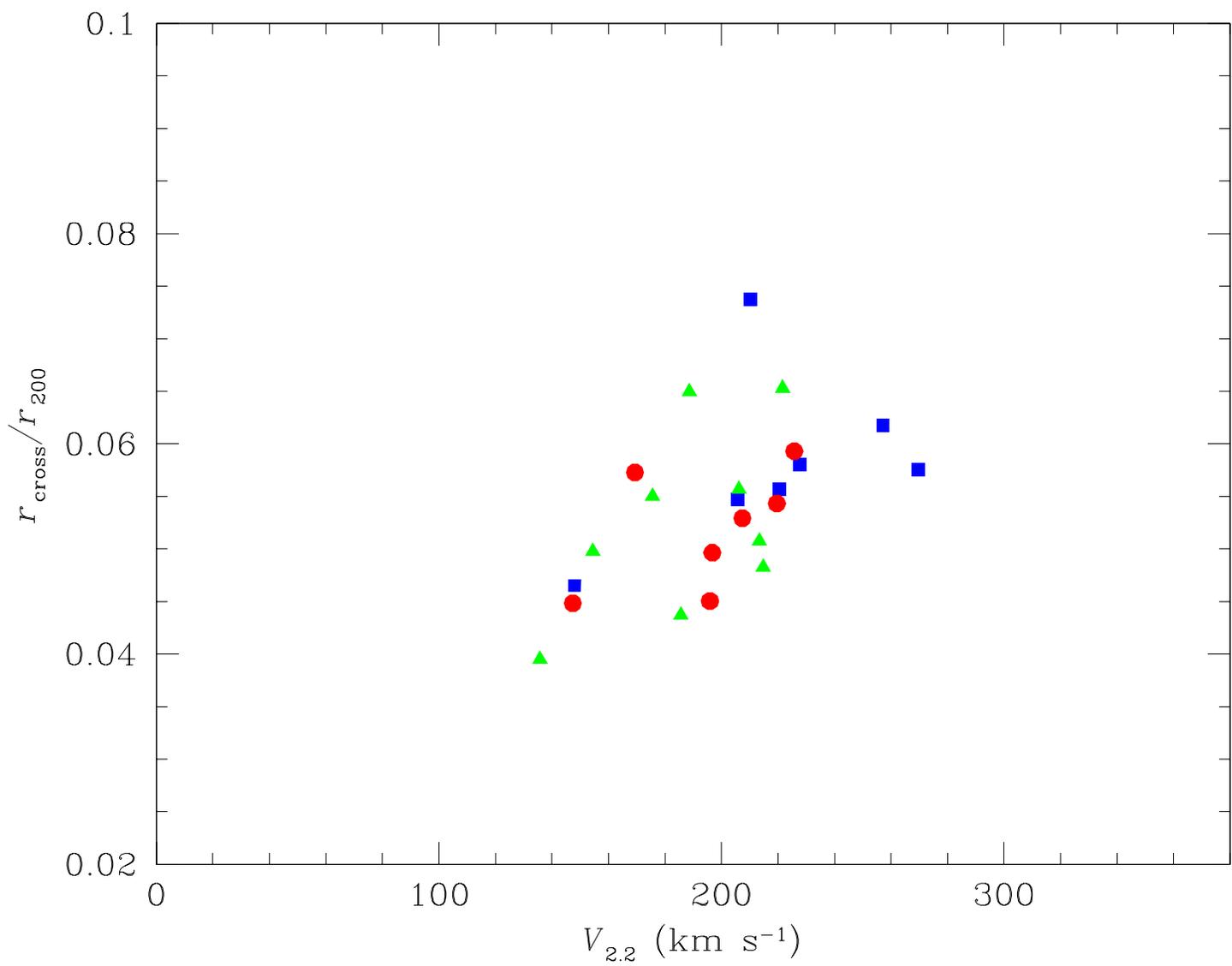
\begin{figure}
  \vspace{2cm}

  \caption{
    The $r_{\rm cross}$ parameter
    divided by the virial radius $r_{200}$ versus the
    $V_{2.2}$ velocities for the DLO sample.
    Symbols are as in Fig.~\ref{NOPTBD}.
    DLOs belonging to haloes that include several objects
    (either DLOs, spheroids or irregulars) have not been considered.
    }
  \label{RCROSS}
\end{figure}

To quantify the importance of dark matter (or baryons) in the
inner regions of DLOs we have used the $r_{\rm cross}$ parameter 
(TDT98).
We define $r_{\rm cross}$ as the radius where 
$V_{\rm bar}(r) = V_{\rm dm}(r)$, such that RCs are dynamically 
dominated by baryons for $r < r_{\rm cross}$, 
and dark matter dominated $r > r_{\rm cross}$ (see Table~\ref{Dynamical}).
Note that most DLOs have
$1 \leq \frac{r_{\rm cross}}{R_{\rm d}} \leq 3$
and only a few have $r_{\rm cross} > R_{2.2}$; note also
that there is no correlation between $r_{\rm cross}$ and $R_{\rm d}$.
In Fig.~\ref{RCROSS} we plot $\frac{r_{\rm cross}}{r_{200}}$ versus $V_{2.2}$;
some correlation appears,
in the sense that more massive objects are less dark matter dominated at
their centres. This correlation disappears when
$V_{2.2}$ is normalized to the halo global velocity, $V_{200}$,
indicating that most of the correlation
seen in Fig.~\ref{RCROSS} is provided by the halo total mass.
In fact, as stated above,
less massive haloes are known  to be more centrally concentrated than
more massive ones.

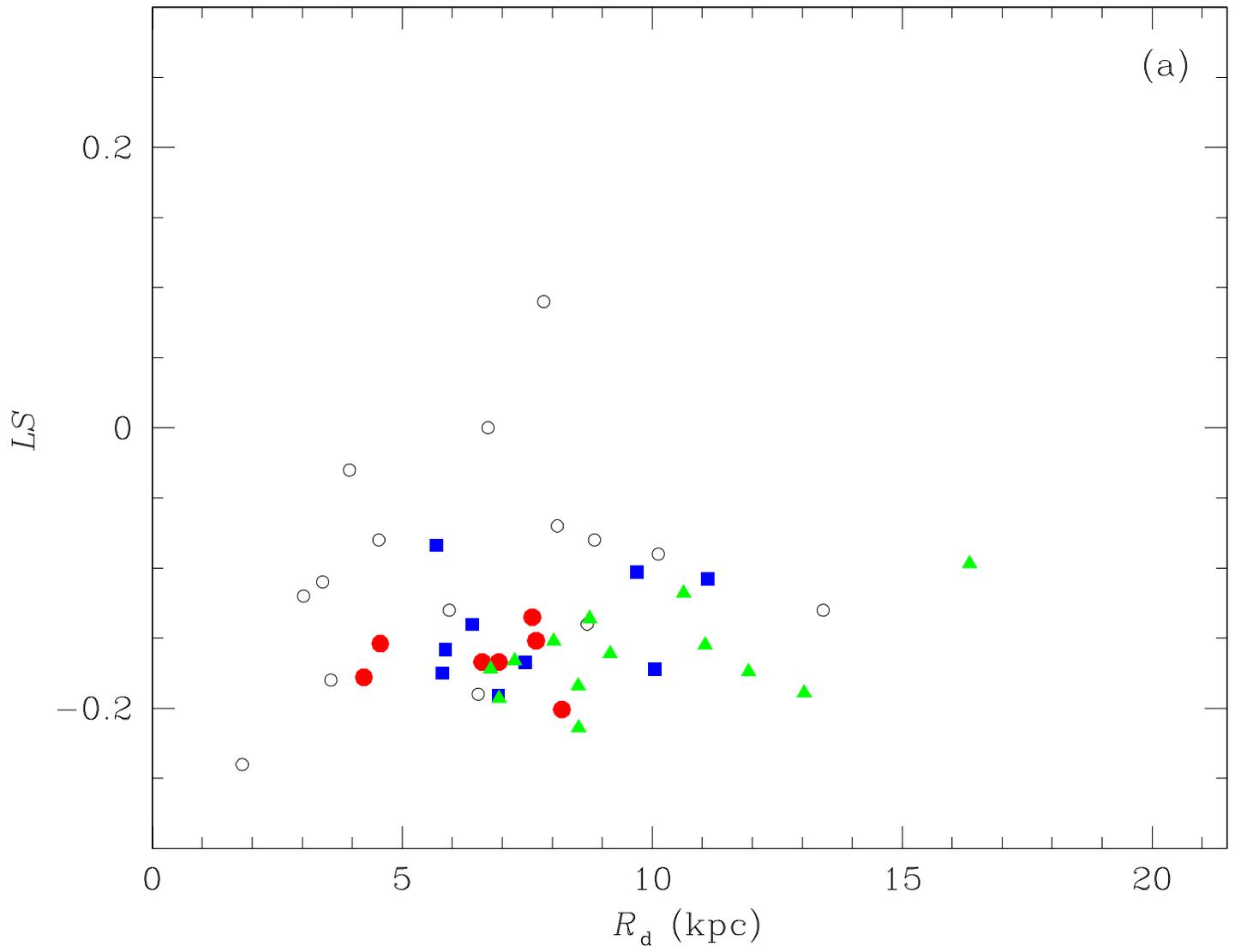
\begin{figure}
  \vspace{2cm}

  \caption{
    The logarithmic slope, $LS$, versus the disc scale lengths $R_{\rm d}$ (a)
    and the $V_{2.2}$ velocities (b) for the DLO sample.
    Squares, triangles and circles correspond
    to DLOs formed in S1, S1b and S1c simulations, respectively.
    Open circles are data from spirals with extended \hbox{H\,{\sc i}} 
    RCs, in the same dynamical range (see table~2 in CvG91).
    }
  \label{LSrd}
\end{figure}

In Figs.~\ref{LSrd}(a)~and~\ref{LSrd}(b), 
we plot $LS$ versus the disc scale lengths $R_{\rm d}$ and
versus  $V_{2.2}$, respectively, for the DLOs and, as a
comparison, the objects of CvG91  compilation in the same dynamical
range. We see that DLOs with $V_{2.2} > 180$ km~s$^{-1}$
do have observational counterparts, i.e., their
$LS$ parameters take values that are found in observed spirals
with the same range in $V_{2.2}$, 
even if we miss  objects
with  $LS$  between, say, $0$ and $-0.1$.
Those  with $V_{2.2} < 180$ km~s$^{-1}$ have  excessively low $LS$s,
but recall that only one galaxy in CvG91
has $V_{2.2} \in \left[150,180\right]$ km~s$^{-1}$. 
Moreover, according to CvG91, $LS$s are correlated with the morphological
type. Our DLOs would be $T \leq 5$, in consistence with the range
in $T$ type previously found from the bulge shape parameters.

It is worthwhile to note that the value of the $LS$ is almost independent
on how baryonic mass is distributed inside $R_{2.2}$, particularly
on how it is shared between the bulge and the disc
(it changes on average by only 10 per cent when the bulge mass is halved 
conserving the baryon number inside $R_{2.2}$). 
The fact that $LS$s are negative may  reflect in part the fact
that dark matter haloes in standard CDM scenarios are too concentrated, leading
to declining rotation curves (Navarro, Frenk \& White 1998). 
In fact, the declining shape of DLO RCs results from 
the flat halo contribution to RCs (see TDT98 for a detailed discussion) 
and the dynamic baryon dominance in the inner DLO regions.
Declining RCs have been observed in spiral galaxies. Early examples
can be found in Bosma (1978, 1981), van~Moorsel (1982).
Then, CvG91, B92 and Bosma (1999), among others, also found other 
examples.  CvG91 suggest that this is a common feature of compact 
bright spiral galaxies, and also frequent in large bright spirals
(although see B92 who shows, contrary to CvG91, that declining
RCs do not correlate with the disc scale length).
Dubinski, Mihos \& Hernquist (1999) have found that the 
declining character of RCs is a necessary condition to form 
long tidal tails in galaxy mergers as those observed in some 
interacting galaxy systems.

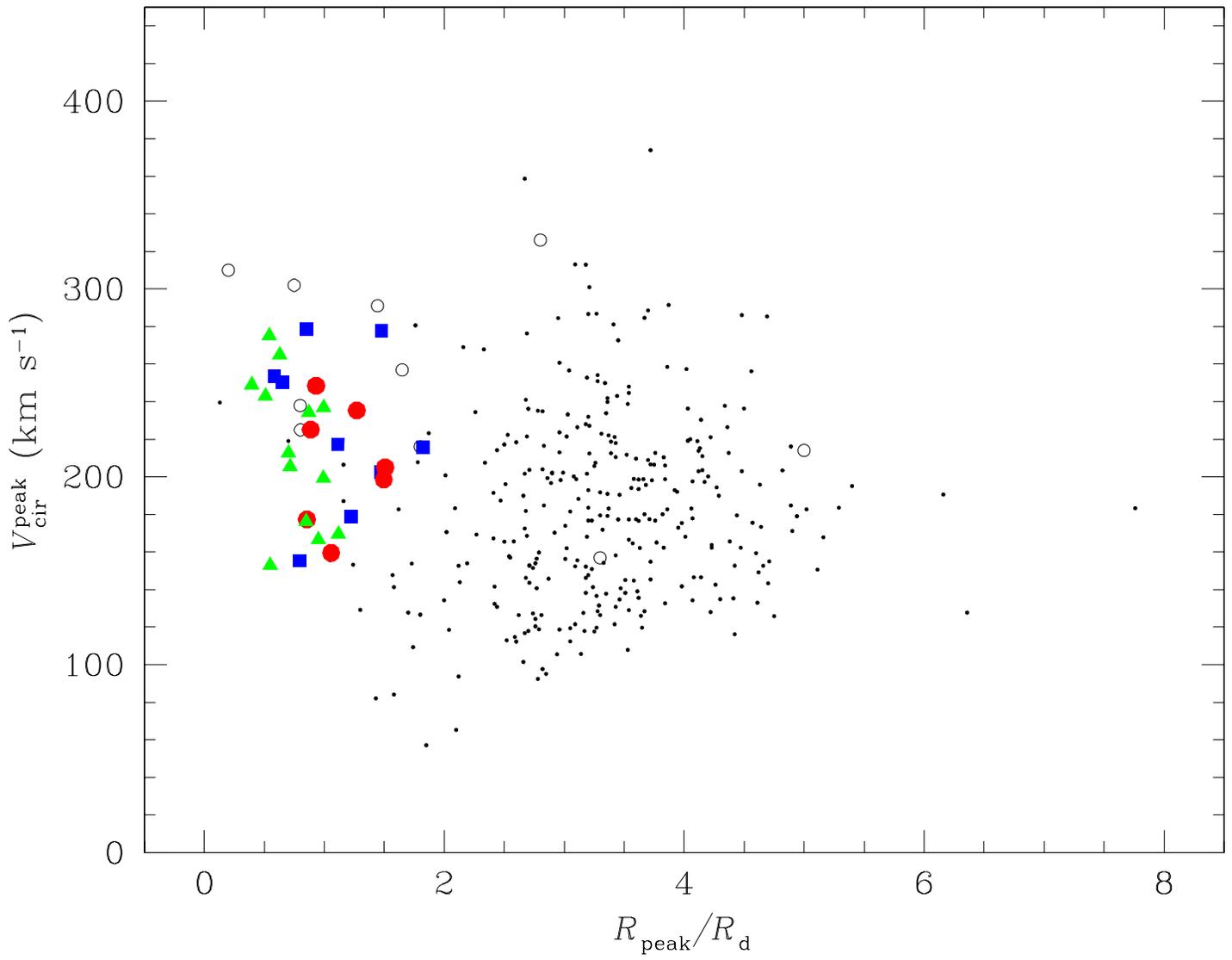
\begin{figure}
  \vspace{2cm}

  \caption{
    The peak circular velocities $V_{\rm cir}^{\rm peak}$ versus the
    $R_{\rm peak}/R_{\rm d}$ ratios for the DLO sample.
    Squares, triangles and circles correspond
    to DLOs formed in S1, S1b and S1c simulations, respectively.
    Dots are data from the Courteau sample of RCs in H$\alpha$;
    open circles are
    data from spirals with extended \hbox{H\,{\sc i}} 
    RCs from table~4.1 in Rhee (1996). See text.
    }
  \label{VRMAX}
\end{figure}

In Fig.~\ref{VRMAX} we plot $V_{\rm cir}^{\rm peak}$ versus the
$R_{\rm peak}/R_{\rm d}$ ratio for the 29 DLOs in our sample.
For comparison, we also give the peak velocities in \hbox{H\,{\sc i}}
of Sb--Sc spirals 
from table~4.1 in RC96 (most of them from B92 compilation),
with $R_{\rm peak}/R_{\rm d}$ ratios taken from C97b. We also
plot the $V_{\rm cir}^{\rm peak}$ and $R_{\rm peak}/R_{\rm d}$ ratios
measured in the Courteau sample of optical RCs (C97b). We confirm the previous
finding that DLOs in Table~\ref{General} have observational counterparts, even
if the smaller ones are somewhat too concentrated as compared with 
Courteau data. Note, however, that $R_{\rm peak}$ is difficult
to measure  in DLOs and 
the measurement hard to assess in  spiral galaxies.

\section{Summary and Conclusions}
\label{sumcon}
We present the results of a detailed and careful comparison
between the parameters characterizing the structural and dynamical
properties of  a sample of 29 simulated DLOs and  those measured
in observed spiral galaxies. These DLOs have been identified in
three fully consistent hierarchical hydrodynamical simulations,
where an {\em inefficient\/} Schmidt law-like algorithm to model the
stellar formation process has been implemented.
In this paper we have been only concerned with disc {\em structural\/} and
{\em dynamical\/} properties, because they keep the imprints of the
dynamical and hydrodynamical processes that, together
with their interplay with star
formation, are  key for galaxy formation and evolution.

The comparison of DLOs in our sample with spiral
discs has been a two step work.
First, DLO bulge-disc decomposition has been performed.
This yields the bulge effective scale length and shape
parameters, $R_{\rm b, eff}$ and $n_{\rm opt}$, the disc scale length,
$R_{\rm d}$, and the mass bulge-to-disc ratio, $B/D$, among
other DLO parameters.
After having tested the robustness of the
fitting procedure, the following results have been obtained:
\begin{enumerate}
\item
  The scale lengths, $R_{\rm b, eff}$ and $R_{\rm d}$,
  and their ratio $R_{\rm b, eff}/R_{\rm d}$, 
  are consistent with available data
  (CdJB96; de Jong 1996; MGH99).
\item
  The distribution of bulge shape parameters is similar to that
  found by CdJB96 for their sample of Sa--Sc galaxies.
\item
  The mass $B/D$ ratios are somewhat high as compared to
  luminosity $B/D$ ratios, but current observational uncertainties on the
  bulge and disc mass-to-light ratios make it difficult to
  draw any conclusion from this comparison.
\end{enumerate}

In a second step, DLO RCs have been analyzed.
DLO gas particles placed at 
distances between 2~kpc and 30~kpc to the centre
move roughly along circular orbits on the equatorial plane.
They are in centrifugal supported equilibrium within
the potential well produced by the total mass distribution
(TDT98). 
Rotation curve shapes have been parametrized through $R_{\rm d}$
(or $R_{2.2} = 2.2 \, R_{\rm d}$) as
spatial scale, $V_{2.2} = V_{\rm cir} (R_{2.2})$ as mass size,
and the maximum or
peak velocity, $V_{\rm cir}^{\rm peak}$, and the radius where
this is reached, $R_{\rm peak}$.
These parameters have been compared with those measured at:
a)~Sb--Sc field spirals drawn from the Courteau sample, with long-slit
H$\alpha$ spectrum and $r$-band photometry (C97b), and
b)~Sb--Sc spirals from Broeils' compilation (B92) of extended
\hbox{H\,{\sc i}} 
RCs with surface photometry. The main results follow:
\begin{enumerate}
\item
  In contrast to findings in other fully-consistent
  hydrodynamical simulations (e.g., Navarro \& Steinmetz 2000;
  Thacker \& Couchman 2000),
  DLO $V_{2.2}$ velocities have been found to be consistent with
  observational data. This is a consequence of disc formation
  with $\bmath j$ conservation.
\item
  The $R_{\rm d}$ and $V_{2.2}$ values obtained indicate that DLOs in our
  sample are either large bright spirals (18), intermediate spirals (9), or
  compact bright spirals (2).
\item
  The average relative contribution of baryons to $V_{2.2}$ in our sample is
  $\left( {V_{\rm bar} \over V_{\rm cir}} \right)_{R_{2.2}} =
  0.67 \pm 0.07$, in very good agreement with Bottema (1993, 1997)
  and CR99, if we take $V_{\rm bar} =  V_{\rm lum}$.
  Most DLOs have been found to have sub-maximal discs;
  only two of them have been found to lie inside the 95 per cent 
  confidence interval for maximum discs according to Sackett (1997).
  The previous agreement also shows that the amount of
  baryon mass that has ended up inside $R < R_{2.2}$ is not excessive,
  again as a consequence of $\bmath j$ conservation.
\item
  The $\frac{r_{\rm cross}}{r_{200}}$ parameter ratio 
  is a measure of the relative amount of
  dark matter at the inner DLO zones.
  Some correlation between these amounts and the circular velocity at 
  $R_{2.2}$  has been found, in the sense that
  less massive objects tend to be more dark matter dominated in their central
  regions. 
\item
  Concerning the parameters that give a measure of the central
  concentrations, the comparison of the $V_{\rm cir}^{\rm peak}$
  versus $R_{\rm peak} \over R_{\rm d}$
  plots to B92 data confirms again that DLOs have observational
  counterparts, even if the smaller DLOs are somewhat too concentrated
  as compared to Courteau data (C97b).
\item
  The logarithmic slope, $LS$, measures the RC slopes for
  $R > R_{2.2}$, where, as stated, in most cases dark matter
  is already dynamically
  dominant. DLOs with $V_{2.2} > 180$ km~s$^{-1}$ occupy the same
  zone in the $LS$ versus $V_{2.2}$ and versus $R_{2.2}$ plots
  as galaxies studied by CvG91 in the same dynamical range;
  those with $V_{2.2} < 180$ km~s$^{-1}$  do not show
  the tendency of less massive galaxies in CvG91 to have higher $LS$.
  This may be a consequence of the strongly concentrated mass
  distribution of dark haloes in standard CDM scenarios.
\end{enumerate}

A concern is in order regarding numerical resolution.
DLOs are resolved with a 
relatively low number of particles. In contrast, dark matter haloes are
described with a much better resolution. An inappropriate low gas resolution
would result in an unphysical gas heating that could halt the gas collapse
(Navarro \& Steinmetz 1997). However, some works suggest that it
is an inadequate resolution in the dark matter halo component that may produce
the larger undesired numerical artifacts (Steinmetz \& White 1997).
In fact, it appears that a well-resolved dark matter halo, even if
the number of gas particles is lower
(but more than, say, one hundred),
gives rise to a well-represented (or
well-resolved) gas density profile, being this point the most important for
both the hydrodynamics and the tracking of the star formation history.
Moreover, to make sure that the populated and extended discs in the
simulations do not result from unphysical gas heating
or smoothing, we have run a higher
resolution simulation ($64^3$ particles in a periodic box of 5~Mpc, with
cosmological and star formation parameters similar to those in S1,
S1b and S1c; hereafter
HRS). Only one disc with mass comparable to those in these lower
resolution simulations forms. It
has 1713+1380 gas+star particles and its halo 36112
dark matter particles. Its analysis
has shown that it is populated and extended, that its structural and dynamical
characteristics are also compatible with observations (see table~1 and fig.~1
in Dom\'{\i}nguez-Tenreiro et al.\ 1998, where we show that the
$R_{\rm b}$ and $R_{\rm d}$
parameters and their ratio take a typical value, and the specific
angular momentum has been conserved) and that the physical processes leading
to its formation are essentially the same as those that are at work in S1, S1b
or S1c.

In conclusion, the comparison between DLOs produced in our simulations and
observational data allows us to affirm that they have counterparts
in the real world.
This agreement suggests that the process
operating in Fall \& Efstathiou (1980) standard
model for disc formation (i.e., gas cooling and collapse with
specific angular momentum, $\bmath j$,  conservation) is also at work in the
{\em quiescent\/} phases of DLO formation in these simulations,
resulting in discs with exponential density profiles,
as predicted by Dalcanton et al.\ (1997). 
However, in Dom\'{\i}nguez-Tenreiro et al.\ (1998) 
and S\'aiz et al.\ 
(1999) it is shown that
{\em violent\/} episodes (i.e., interaction and merger events) also occur
and play an important role in DLO assembly. In particular, it is
shown that dark matter haloes formed in these simulations are not
always able to stabilize the {\em pure\/} exponential disc that
would form at its central region
according to Fall \& Efstathiou and Dalcanton  et al.\ 
scenarios (see also van den
Bosch 2000).
To provide the right conditions for disc regeneration to occur
after the last violent episode of DLO assembly, a compact central
bulge is needed; this will ensure the axisymmetric character
of the gravitational potential well at scales of some kpcs at all
times, avoiding excessive $\bmath j$ losses in violent events.
In TSDT these arguments are developed, and, moreover,
we prove that a second condition is necessary for disc
regeneration: the availability of gas at low~$z$, that is,
it is necessary that
the SF algorithm implementation does not result
in a too early gas exhaustion.
The good behaviour of our DLO sample as compared with
observations suggests that the {\em inefficient\/} SF algorithm used
in the simulations
has met both requirements.

The global agreement  with observations we have found,
at a structural and dynamical level, also represents
an important step towards making numerical approaches more widely used
to learn about galaxy formation and evolution
in a cosmological framework, i.e., from primordial fluctuations.
These approaches will be
particularly useful if, as expected, future improvements in
numerical methods and computers speed allow to numerically
simulate the different aspects of galaxy assembly with an increasingly
high degree of realism.

\section*{Acknowledgments}

It is a pleasure to thank  E. Bertschinger for providing us with some
software subroutines, J.~F. Navarro for interesting discussions
and J. Silk for his valuable comments on our work.
We wish to acknowledge as well the referee, J. Sommer-Larsen, for his
suggestions and interesting comments.
This work was supported in part by DGES (Spain) through grants
number PB93-0252 and PB96-0029. AS was also supported
by DGES through fellowships.
PBT thanks DGES, and CONICET and ANPCyT (Argentina) for their
financial support.
We are indebted to the Centro de Computaci\'on Cient\'{\i}fica (Universidad
Aut\'onoma de Madrid) and to the Oxford University
for providing the computational support to perform
this work.

\clearpage

\begin{table}
  \caption{General Parameters}
  \label{General}
  \begin{tabular}{lllrrrrrlrrr}
    &&&&& \multicolumn{3}{c}{$r_{200}$} && \multicolumn{3}{c}{$R_{\rm max}$}\\
    \cline{6-8} \cline{10-12}
    Simulation& DLO\# & $r_{200}$& $R_{\rm max}$ &&
    $N_{\rm dark}$&$N_{\rm gas}$&$N_{\rm star}$&&
    $N_{\rm dark}$&$N_{\rm gas}$&$N_{\rm star}$\\
    \hline
    S1  &126   &239.  &25.9 &&  2709 & 248 & 157 &&  516 &185 &157\\
        &143   &194.  &37.5 &&  1482 & 153 &  47 &&  473 &110 & 47\\
        &233A  &310.* &28.5 &(& 6027 & 583 & 164 &)& 506 &201 & 95\\
        &233B  &310.* &25.4 &(& 6027 & 583 & 164 &)& 400 &148 & 69\\
        &242   &316.  &33.4 &&  6325 & 621 & 281 &&  959 &339 &278\\
        &333   &296.  &49.7 &&  5213 & 509 & 217 && 1337 &338 &212\\
        &531   &235.  &30.9 &&  2573 & 275 &  79 &&  589 &224 & 79\\
        &544   &312.  &43.4 &&  6170 & 555 & 230 && 1304 &349 &215\\
        &545   &328.  &26.2 &&  7131 & 671 & 252 &&  794 &381 &238\\
    \\
    S1b &054   &252.  &35.9 &&  3076 & 299 & 138 &&  611 &171 &138\\
        &110   &254.  &30.3 &&  3145 & 268 & 174 &&  648 &163 &174\\
        &116$\alpha$  &303.  &73.0 &&  5357 & 455 & 270 && 1806 &291 &269\\
        &116$\beta$  &178.  &38.1 &&  1029 & 171 &  30 &&  460 &138 & 30\\
        &164   &222.  &29.2 &&  2107 & 233 &  62 &&  496 &160 & 62\\
        &165C  &467.* &39.1 &(&19522 &1188 &1578 &)& 556 &113 & 56\\
        &211   &186.  &43.5 &&  1236 & 157 &  12 &&  468 &130 & 12\\
        &212   &261.  &49.4 &&  3419 & 226 & 237 && 1171 &192 &234\\
        &633A   &312.* &40.9 &(& 5890 & 371 & 401 &)&1068 &188 &325\\
        &643$\alpha$  &317.  &47.4 &&  6176 & 447 & 373 && 1217 &180 &358\\
        &643$\beta$A  &307.* &38.1 &(& 5583 & 518 & 270 &)& 695 &217 &213\\
        &643$\beta$B  &307.* &32.4 &(& 5583 & 518 & 270 &)& 351 &146 & 55\\
        &643$\gamma$  &232.  &40.8 &&  2409 & 234 &  97 &&  683 &181 & 97\\
    \\
    S1c &106   &217.  &29.5 &&  1966 & 196 &  76 &&  419 &122 & 76\\
        &324   &191.  &36.6 &&  1322 & 173 &  22 &&  395 &127 & 22\\
        &342   &280.  &34.3 &&  4214 & 470 & 114 &&  770 &311 &107\\
        &444  &267.  &33.8 &&  3727 & 352 &  99 &&  737 &243 & 99\\
        &454  &215.  &20.4 &&  1907 & 184 &  70 &&  383 &149 & 70\\
        &556   &265.  &31.0 &&  3605 & 257 & 229 &&  707 &178 &229\\
        &631   &215.  &18.9 &&  1917 & 193 &  59 &&  322 &145 & 59\\
    \\

  \end{tabular}
  
  \medskip
  Distances are given in kpc. An asterisk
  indicates that the corresponding  DLO is embedded 
  in a double or multiple halo,
  i.e., that one or more other baryonic objects (either disc-like or not)
  exist within $r_{200}$. This is to be taken into account
  on evaluating the total number of halo particles  (numbers in
  parentheses).
\end{table}

\clearpage

\begin{table}
  \caption{Bulge/disc parameters}
  \label{BulDis}
  \begin{tabular}{lccrcrclcrcrclrr}
    &\multicolumn{6}{c}{Optimal Fit}&& \multicolumn{5}{c}{$n=1$}&& \multicolumn{2}{c
      }{$n=4$}\\
    \cline{2-7} \cline{9-13} \cline{15-16}
    DLO\# &  $n_{\rm opt}$&  $R_{\rm b,eff}$& $R_{\rm d}$& $B/D$& $\chi^2/N$& $R_{\rm trans}$ &&
    $R_{\rm b,eff}$& $R_{\rm d}$& $B/D$& $\chi^2/N$& $R_{\rm trans}$ &&
    $R_{\rm d}$& $\chi^2/N$\\ 
    \hline
    
    126  &2.6 &0.90 & 6.3 &1.83 & 5.72 &4.46 &&1.09 & 5.8 &1.48 &10.40 &3.50 && 6.5 & 6.35\\
    143  &3.2 &0.58 &10.9 &1.45 & 1.61 &4.54 &&0.91 &10.1 &1.36 & 2.30 &3.52 &&11.1 & 1.53\\
    233A &1.3 &1.00 & 6.5 &1.25 & 5.52 &3.57 &&1.08 & 6.4 &1.22 & 5.55 &3.43 && 7.1 & 5.81\\
    233B &0.6 &1.05 & 5.6 &1.22 & 9.09 &2.77 &&0.89 & 5.7 &1.23 & 9.55 &2.89 && 5.8 &12.54\\
    242  &1.2 &1.25 & 7.5 &1.23 &15.96 &4.19 &&1.29 & 7.5 &1.21 &16.19 &4.06 && 7.9 &42.04\\
    333  &1.9 &1.36 &12.0 &1.72 & 5.30 &6.59 &&1.50 &11.1 &1.58 &11.40 &5.36 &&12.8 &11.98\\
    531  &0.9 &1.70 & 6.9 &2.02 & 3.85 &5.24 &&1.69 & 6.9 &2.07 & 4.00 &5.39 && 9.3 &19.57\\
    544  &2.3 &1.05 &10.6 &1.79 & 2.44 &5.82 &&1.26 & 9.7 &1.61 & 9.05 &4.54 &&11.3 & 4.06\\
    545  &1.1 &1.26 & 5.9 &1.23 & 3.01 &3.78 &&1.27 & 5.9 &1.22 & 3.11 &3.74 && 6.4 &20.24\\
    \\
    054  &1.6 &0.89 & 8.4 &2.99 & 2.94 &4.80 &&1.04 & 8.0 &2.82 & 3.65 &4.13 && 9.7 & 4.32\\
    110  &1.8 &0.60 & 7.0 &2.54 & 2.64 &3.46 &&0.78 & 6.8 &2.47 & 3.28 &3.13 && 7.2 & 3.46\\
    116$\alpha$ &2.3 &0.48 &16.4 &1.40 &13.97 &4.12 &&0.76 &16.3 &1.39 &14.55 &3.50 &&16.5 &14.79\\
    116$\beta$ &1.3 &1.26 & 8.7 &1.70 & 1.75 &4.97 &&1.34 & 8.5 &1.64 & 1.93 &4.60 &&10.1 & 3.10\\
    164  &1.3 &1.19 & 7.2 &2.47 & 1.70 &4.83 &&1.25 & 6.9 &2.36 & 1.99 &4.44 && 9.9 & 4.64\\
    165C &1.6 &0.93 & 9.0 &2.37 & 1.08 &4.67 &&1.06 & 8.7 &2.28 & 1.28 &4.15 && 9.8 & 1.71\\
    211  &1.8 &1.35 &14.0 &1.42 & 0.75 &6.47 &&1.49 &13.0 &1.36 & 1.19 &5.40 &&17.0 & 1.30\\
    212  &2.3 &0.53 &11.2 &1.95 & 4.93 &4.04 &&0.80 &11.1 &1.91 & 6.30 &3.46 &&11.3 & 5.49\\
    633A  &1.1 &0.72 & 9.0 &3.01 & 4.69 &3.33 &&0.74 & 9.2 &2.98 & 4.33 &3.32 && 9.6 & 6.62\\
    643$\alpha$ &1.4 &1.19 &10.9 &3.40 & 3.14 &5.91 &&1.28 &10.6 &3.27 & 5.07 &5.30 &&13.2 &19.61\\
    643$\beta$A &2.8 &0.86 & 9.4 &2.36 & 3.18 &5.75 &&1.13 & 8.5 &2.02 &10.91 &4.22 &&10.0 & 3.70\\
    643$\beta$B &1.4 &1.28 & 7.6 &1.98 & 1.46 &5.02 &&1.35 & 7.3 &1.87 & 1.96 &4.54 && 9.5 & 3.45\\
    643$\gamma$ &1.7 &1.18 &13.2 &2.43 & 2.90 &6.56 &&1.32 &11.9 &2.32 & 4.33 &5.30 &&17.8 & 5.37\\
    \\
    106  &2.0 &1.06 & 7.5 &2.69 & 1.18 &5.51 &&1.20 & 6.6 &2.25 & 2.47 &4.22 && 8.2 & 1.70\\
    324  &1.2 &1.42 & 8.3 &1.96 & 0.80 &5.22 &&1.46 & 8.2 &1.91 & 0.84 &4.99 &&10.1 & 2.56\\
    342  &1.3 &1.43 & 8.2 &1.45 & 2.52 &5.09 &&1.46 & 7.7 &1.38 & 2.87 &4.60 && 9.2 & 9.95\\
    444 &1.3 &1.41 & 7.9 &2.02 & 3.49 &5.34 &&1.46 & 7.6 &1.92 & 4.22 &4.89 &&10.9 &11.14\\
    454 &1.0 &1.23 & 4.6 &3.14 & 2.33 &4.26 &&1.24 & 4.6 &3.10 & 2.34 &4.20 &&10.2 & 6.45\\
    556  &2.1 &0.86 & 7.6 &3.70 & 1.64 &5.37 &&1.06 & 6.9 &3.17 & 6.17 &4.13 && 8.2 & 3.70\\
    631  &0.9 &1.13 & 4.1 &2.36 & 0.97 &3.50 &&1.11 & 4.2 &2.43 & 0.97 &3.62 &&18.0 & 2.78\\

  \end{tabular}

  \medskip
  Distances are given in kpc.
\end{table}

\clearpage

\begin{table}
  \caption{Dynamical Parameters}
  \label{Dynamical}
  \begin{tabular}{llllccccccccc}
    &&&&&&&& \multicolumn{2}{c}{$n=1$} && \multicolumn{2}{c}{Optimal Fit}\\
    \cline{9-10} \cline{12-13}
    DLO\# & $V_{200}$& \multicolumn{1}{c}{$\lambda$}&  $r_{\rm cross}$& $LS$ & $R_{\rm peak}$& $V_{\rm cir}^{\rm peak}$ &&$V_{2.2}$ & $V_{\rm bar,2.2}$&&$V_{2.2}$ & $V_{\rm bar,2.2}$\\
    \hline
    126  & 114.  & 0.039  & 16.6 & -0.176 & 10.6 & 216. && 210. & 157. && 206.  & 153.\\
    143  & 98.5  & 0.045  & 8.94 & -0.172 & 7.98 & 156. && 148. & 83.0 && 147.  & 80.6\\
    233A & 157.* & 0.063* & 14.0 & -0.139 & 9.45 & 203. && 197. & 139. && 197.  & 139.\\
    233B & 157.* & 0.063* & 12.9 & -0.085 & 6.93 & 179. && 176. & 125. && 176.  & 125.\\
    242  & 160.  & 0.069  & 19.3 & -0.168 & 6.33 & 279. && 257. & 186. && 257. & 186.\\
    333  & 150.  & 0.079  & 16.2 & -0.108 & 7.25 & 250. && 221. & 143. && 218. & 139.\\
    531  & 118.  & 0.011  & 12.7 & -0.192 & 7.70 & 217. && 206. & 140. && 206. & 140.\\
    544  & 158.  & 0.023  & 17.8 & -0.104 & 5.69 & 254. && 228. & 153. && 225. & 148.\\
    545  & 166.  & 0.049  & 18.6 & -0.159 & 8.66 & 278. && 270. & 204. && 269. & 204.\\
    \\
    054  & 126.  & 0.122  & 16.4 & -0.152 & 5.64 & 213. && 189. & 132. && 187. & 129.\\
    110  & 127.  & 0.044  & 12.2 & -0.172 & 6.74 & 237. && 215. & 148. && 214. & 147.\\
    116$\alpha$ & 151.  & 0.067  & 16.8 & -0.097 & 6.46 & 249. && 206. & 121. && 206. & 121.\\
    116$\beta$ & 88.8  & 0.055  & 8.80 & -0.214 & 8.11 & 167. && 155. & 92.6 && 153. & 92.0\\
    164  & 111.  & 0.038  & 9.63 & -0.193 & 6.88 & 199. && 186. & 120. && 184. & 118.\\
    165C & 228.* & 0.043* & 7.70 & -0.136 & 7.43 & 176. && 160. & 92.7 && 160. & 92.6\\
    211  & 92.7  & 0.095  & 7.29 & -0.189 & 7.15 & 153. && 136. & 70.8 && 135. & 69.0\\
    212  & 130.  & 0.062  & 13.2 & -0.155 & 5.64 & 243. && 213. & 132. && 213. & 131.\\
    633A  & 156.* & 0.067* & 19.9 & -0.161 & 5.78 & 265. && 227. & 160. && 227. & 161.\\
    643$\alpha$ & 158.  & 0.086  & 20.6 & -0.118 & 5.78 & 275. && 222. & 153. && 222. & 152.\\
    643$\beta$A & 154.* & 0.067* & 20.3 & -0.184 & 7.43 & 234. && 209. & 149. && 204. & 144.\\
    643$\beta$B & 154.* & 0.067* & 15.1 & -0.166 & 8.11 & 170. && 157. & 110. && 156. & 108.\\
    643$\gamma$ & 116.  & 0.054  & 12.7 & -0.174 & 8.53 & 205. && 176. & 105. && 171. & 100.\\
    \\
    106  & 108.  & 0.075  & 12.4 & -0.167 & 5.64 & 177. && 169. & 116. && 167. & 110.\\
    324  & 95.2  & 0.045  & 8.53 & -0.201 & 8.66 & 160. && 147. & 89.9 && 147. & 88.8\\
    342  & 140.  & 0.018  & 15.1 & -0.152 & 9.76 & 235. && 220. & 153. && 217. & 149.\\
    444 & 133.  & 0.055  & 14.0 & -0.135 & 6.74 & 225. && 207. & 141. && 207. & 139.\\
    454 & 107.  & 0.034  & 10.6 & -0.154 & 6.88 & 205. && 197. & 141. && 197. & 141.\\
    556  & 130.  & 0.063  & 15.4 & -0.167 & 6.46 & 248. && 226. & 160. && 222. & 155.\\
    631  & 107.  & 0.025  & 9.63 & -0.178 & 6.33 & 199. && 196. & 140. && 196. & 141.\\
    \\
  \end{tabular}

  \medskip
  Distances are given in kpc; velocities in km~s$^{-1}$.
  Asterisks stand for double or multiple haloes, so that their global
  parameters (with an asterisk in this Table) might not reflect
  a direct property of the corresponding DLO.
\end{table}

\clearpage

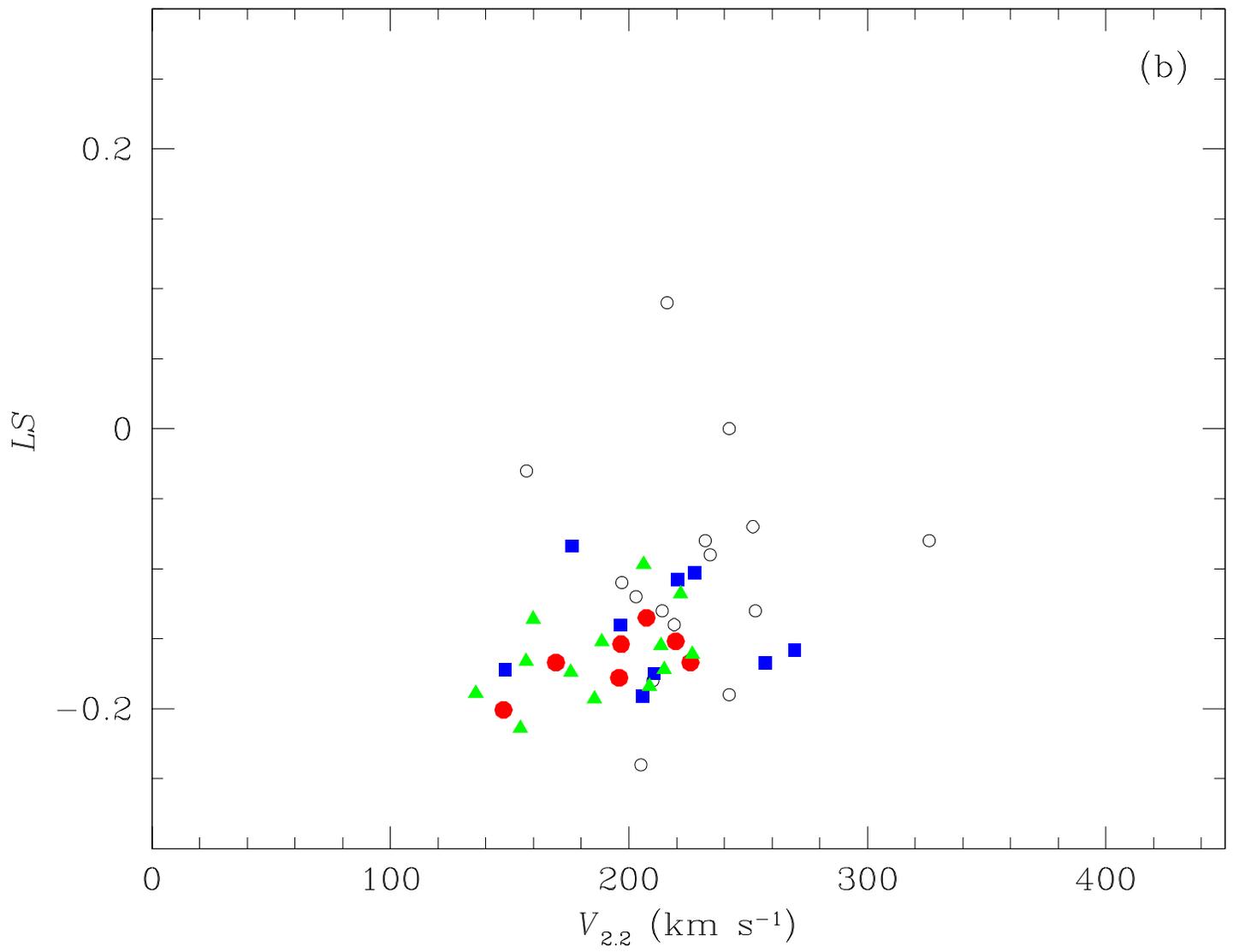
\end{document}